\documentclass[a4paper]{article}

\usepackage{multirow}
\usepackage{amsmath,amsfonts,amssymb}
\usepackage{amsthm}
\usepackage{graphicx}
\usepackage{caption}
\usepackage{subcaption}

\usepackage{hyperref}

\usepackage[ruled,vlined,linesnumbered]{algorithm2e}
\usepackage{appendix}
\usepackage{siunitx}

\usepackage{xcolor}
\usepackage{comment}

\usepackage{authblk}

\usepackage{geometry}
\usepackage{graphicx,float,tabularx}
\usepackage{tabularray} % for \hline[dashed]
\usepackage{xcolor}
\usepackage{xurl}

\newtheorem{definition}{Definition}[section]

\def\BC{{\mathtt{B}}} % blockchain
\def\DSB{{\mathcal{B}}} % benchmark dataset
\def\DSW{{\mathcal{W}}} % whole dataset
\def\DSA{{\mathcal{A}}} % attack dataset
\def\DS{{\mathcal{D}}} % general dataset
\def\SP{{\mathtt{SP}}} % sentiment profile
\def\ei{{\mathtt{ei}}} % emotion profile of a data unit
\def\ev{{\mathtt{ev}}} % emotion profile of a data unit
\def\EI{{\mathtt{EI}}} % emotion intensity of a data set
\def\EV{{\mathtt{EV}}} % emotion volume of a data set
\def\EP{{\mathtt{EP}}} % emotion profile of a data set

\date{}

\begin{document}

\title{Social Media Perceptions of 51\% Attacks on Proof-of-Work Cryptocurrencies: A Natural Language Processing Approach}

% -----------------------------------------
% Sanjay - commented the following out for double blinded review process
\author[1]{Zsofia Baruwa\thanks{All authors contributed equally to this research.}}
%  \authornote{All authors contributed equally to this research.}
%  \email{zb78@kent.ac.uk}
%  \affiliation{%
%    \institution{University of Kent}
%    %\streetaddress{}
%    %\city{Dublin}
%    %\state{Ohio}
%    \country{United Kingdom}
%    %\postcode{43017-6221}
%  }
% 
\author[1]{Sanjay Bhattacherjee\thanks{Corresponding author.}}
%  \authornote{Corresponding author.}
%  \email{s.bhattacherjee@kent.ac.uk}
%  \affiliation{%
%    \institution{University of Kent}
%    %\streetaddress{}
%    %\city{Dublin}
%    %\state{Ohio}
%    \country{United Kingdom}
%    %\postcode{43017-6221}
%  }
% 
\author[2]{Sahil Rey Chandnani}
%  \email{sahilreycc@gmail.com}
%  \affiliation{%
%    \institution{University of Kent}
%    %\streetaddress{}
%    %\city{Dublin}
%    %\state{Ohio}
%    \country{United Kingdom}
%    %\postcode{43017-6221}
%  }
% 
\author[1]{Zhen Zhu}
%  \email{z.zhu@kent.ac.uk}
%  \affiliation{%
%    \institution{University of Kent}
%    %\streetaddress{}
%    %\city{Dublin}
%    %\state{Ohio}
%    \country{United Kingdom}
%    %\postcode{43017-6221}
%  }
% 
\affil[1]{University of Kent}
\affil[ ]{\textit {zb78@kent.ac.uk,s.bhattacherjee@kent.ac.uk,sahilreycc@gmail.com,z.zhu@kent.ac.uk}}
\affil[2]{}
% -----------------------------------------

\maketitle

\begin{abstract}
This work is the first study on the effects of attacks on cryptocurrencies as expressed in the sentiments and emotions of social media users.
Our goals are to design the methodologies for the study including data collection, conduct volumetric and temporal analyses of the data, and profile the sentiments and emotions that emerge from the data.
As a first step, we have created a first-of-its-kind comprehensive list of $31$ events of $51\%$ attacks on various PoW cryptocurrencies, showing that these events are quite common contrary to the general perception.
We have gathered Twitter data on the events as well as benchmark data during normal times for comparison.
We have defined parameters for profiling the datasets based on their sentiments and emotions.
We have studied the variation of these sentiment and emotion profiles when a cryptocurrency is under attack and the benchmark otherwise, between multiple attack events of the same cryptocurrency, and between different cryptocurrencies.
Our results confirm some expected overall behaviour and reactions while providing nuanced insights that may not be obvious or may even be considered surprising.
Our code and datasets are publicly accessible.

% Word count: 185
\end{abstract}

\paragraph{Keywords:} Blockchain, cryptocurrency, $51\%$ attack, sentiment analysis, emotion detection, NLP.

\section{Introduction\label{sec:intro}}

Bitcoin~\cite{Nak08}, the first cryptocurrency was deployed in 2009.
Since then, there has been a flurry of new cryptocurrencies that are in one way or the other, adaptations of the blockchain protocol underlying Bitcoin.
Thousands of cryptocurrencies have come up since then~\cite{statista2022} with a total market capitalisation of more than $1.6$ trillion US dollars\footnote{As on \today~according to~\url{https://coinmarketcap.com/}}.

A unit of the cryptocurrency is called a \textit{coin}.
The blockchain data structure underlying these cryptocurrencies is a hashed chain of blocks that works as a ledger of transactions for the creation and transfer of its coins.
To use these coins as a currency, it is necessary that one should not be able to spend the same coin more than once.
The ledger is public and is expected to provide an \textit{immutable history} of the transactions, thus preventing multiple spendings of the same coin.

\textit{The most important attack on a cryptocurrency is called the $51\%$ attack that defeats this fundamental immutability requirement of its underlying blockchain}~\cite{Nak08,NBFMG16,ES14,GKL15,BS2023}.
Preventing this attack is the primary security goal of a blockchain, as described in Nakamoto's whitepaper~\cite{Nak08} introducing Bitcoin and blockchains to the world.
It allows the attacker to rewrite the transaction history recorded in the ledger.
As a result, transaction(s) denoting transfer of coin(s) are removed from the history, allowing the coins to be spent again.
Hence it is also called the \textit{double-spending} attack.
Other attacks on blockchains have emerged eventually as the blockchain ecosystem has grown and evolved, but the $51\%$ attack remains as the fundamental security goal of a blockchain.
A brief description of the $51\%$ attack is provided in Appendix~\ref{sec:forks-51attacks}.

Cryptocurrencies are decentralised systems maintained by entities called \textit{miners} in proof-of-work (PoW) systems.
Miners contribute new blocks to the underlying blockchain\footnote{In proof-of-stake systems, these entities are the validators/stakers. However, the cryptocurrencies considered in this work are all proof-of-work based systems maintained by miners. So we call these entities miners in general.}.
A $51\%$ attack is conducted by (subsets of) malicious miners controlling more than half of the total mining resources invested in the cryptocurrency system at that point in time\footnote{This is assuming that the network is fully connected. Double spending attacks can be conducted by controlling a smaller percentage of the network hash-rate and splitting the network~\cite{ES14}.}.
Hence the name $51\%$ attack.
% \textcolor{red}{Both reviewers: Why $51\%$ attacks and why not others}

There is a clear \textit{interdependence} between the market value of a cryptocurrency and its security.
As the value of a cryptocurrency increases, more people ``invest'' to profit from it, and it becomes more difficult to launch a $51\%$ attack on it.
This is easy to see in the case of proof-of-work cryptocurrencies.
For example, in Bitcoin, as its market value grows, the miners generally want to increase their computational powers so that their chances of mining new blocks successfully increases and consequently they earn more.
As the total computational power invested in the system increases, it becomes more difficult to mine new blocks.
At the same time, attacking the system also becomes more difficult.
On the other hand, as the computational power of the network decreases -- perhaps due to devaluation of the cryptocurrency -- the miners move to mine other more profitable cryptocurrencies where they can use the same computational power.
Thus, the difficulty of mining new blocks reduces and hence attacking the system becomes easier~\cite{Attack1Vertcoin2018-2}.

Importantly, the converse is also true.
% as it has been observed for every cryptocurrency.
That is, if a cryptocurrency system is threatened by or actually suffers a $51\%$ attack, its value decreases~\cite{Attack2BitcoinSV2021-2}.
In $2014$, the
% first major alarm
mere possibility 
of a $51\%$ attack on Bitcoin~\cite{AttackBitcoin2014-1,AttackBitcoin2014-2,AttackBitcoin2014-3} resulted in an immediate fall in its price~\cite{51AttackBTC_PCWorld}\footnote{The mining pool {\tt GHash.io} attained more than $51\%$ computational power of the Bitcoin network through its participating miners.
Several miners withdrew from the pool to reduce its computational capability and alleviate the possibility of an attack.}.
% The effect of such an attack on the price of the respective cryptocurrency has been reported several times~\cite{51AttackBTC_Medium}.
In~\cite{SSVK2019}, the authors studied $14$ such attacks and found that in each case the price of the attacked cryptocurrency immediately decreased by $12\%$ to $15\%$.
We note here that it~\cite{SSVK2019} is the only other resource on the internet before our work where a substantial number of $51\%$ attacks have been reported to the best of our knowledge.
They also noted that the price does not recover to the pre-attack level for a week after the event.
% \textcolor{red}{Reviewer 2: "It is not really clear if the $51\%$ attack was the actual reason for the loss of the price in cryptocurrencies, as the whole business model is speculative."}
Even though the cryptocurrency market is significantly speculative with various factors affecting the prices, the above result shows the indisputable severity of the $51\%$ attacks in terms of their effect on the prices.

A crucial factor in determining the value of a cryptocurrency is the trust of its users in the immutability of the ledger and other aspects of its security.
Hence popular cryptocurrency risk analysis systems like the CertiK Skynet~\cite{certik-skynet} use ``community trust'' as a key parameter in the security score of blockchain-based systems.
We point out that the perceptions regarding the security of cryptocurrencies are intermediary between attacks and their effect on the prices, creating the need for security scores like those provided by CertiK\footnote{The security score for Bitcoin is available at: \url{https://skynet.certik.com/projects/bitcoin}.}.
The impact of users' perceptions on the price movements of other financial instruments (but not cryptocurrencies) has long been studied in the literature \cite{das2007yahoo,zimbra2015stakeholder,chen2014wisdom,bukovina2016social,tetlock2014information,loughran2016textual}.
Similarly, user perceptions are a determining factor for cryptocurrency prices~\cite{aslam2022sentiment}, much like other financial instruments.
Previous works~\cite{FMBHP2018,MUMT2020} studying user trust on cryptocurrencies have considered aspects like lack of institutional backing, pseudonymity of users, their legal status, volatility of value, reputation of users, etc.
However, the effect that any kind of attack on a cryptocurrency has on users' perceptions has not been studied before.
In this work, we concentrate on this unexplored problem -- \textit{how do attacks on cryptocurrencies affect users' perceptions of them}?

Being a major social media platform, X (formerly Twitter) has been increasingly influential in generating and portraying perceptions by disseminating information.
Such information and the consequent price volatility of financial instruments affect investment decisions in real time~\cite{yang2015twitter,afanasyev2021strength,gross2019buzzwords,behrendt2018twitter,mohan2019stock}.
Twitter data are primarily text-based and therefore are typically analysed with natural language processing (NLP) methods such as sentiment analysis and emotion detection~\cite{behrendt2018twitter,gross2019buzzwords,garcia2023targeted}.
% \textcolor{red}{Both reviewers: Why sentiment analysis is essential for $51\%$ attacks and why not other attacks, what is the purpose to use it: "make a case why SA is necessary, important or even relevant to security practice".}
Both sentiments and emotions detected from Twitter do correlate with stock market movements~\cite{bollen2011twitter} and such data can be used to predict stock price movements, days ahead~\cite{smailovic2013predictive}.
Whereas sentiment analysis often classifies texts into three polarity levels (i.e. negative, neutral, or positive), emotion detection is focused on more granular multi-class identification of human emotions such as anger, fear, joy, love, sadness, and surprise~\cite{nandwani2021review}.
Given that the $51\%$ attack is the most fundamental security risk for cryptocurrencies, we frame the following questions.
% \red{
\begin{itemize}
    \item How common are $51\%$ attacks on proof-of-work cryptocurrencies?
    \item How much and how quickly does such an attack reflect on social media?
    \item What sentiments and emotions do the users generally have about these cryptocurrencies and how do they vary at the time of a $51\%$ attack?
    \item Do users of all cryptocurrencies demonstrate similar sentiments and emotions?
\end{itemize}
% }

\subsection{Our contributions\label{sec:our-contributions}}

This is the first study to profile the effects of any kind of attack on cryptocurrencies as expressed in the sentiments and emotions of social media users.
Given the importance of the $51\%$ attack, it is the obvious choice for this first study.
To begin with, a list of $51\%$ attack events is required for such a study.
So our first major contribution is to create a comprehensive timeline of $51\%$ attack events, providing cross-references of online sources for most events.
Only one such list of $14$ attack events on $13$ mostly minor coins (by market shares) and just one repeated attack has been reported earlier to the best of our knowledge~\cite{SSVK2019}.
This work analysed the effect of $51\%$ attacks on the prices of the cryptocurrencies.
We have identified a total of $31$ events of $51\%$ attacks on $20$ different cryptocurrencies, since the introduction of Bitcoin in 2009.
Our list reveals that $6$ out of these $20$ cryptocurrencies have been attacked multiple times --
Ethereum Classic ($4$ times),
Bitcoin Gold ($3$ times),
Bitcoin SV ($2$ times),
Litecoin Cash ($2$ times),
Verge ($2$ times) and
Vertcoin ($2$ times)
-- all of which had significant market shares at the time of the attacks.
This is presently the most comprehensive list of such real-life attack events to the best of our knowledge.
It shows that \textit{$51\%$ attack events have been quite common in the cryptocurrency ecosystem}.\footnote{The estimated cost of a $51\%$ attack on various proof-of-work cryptocurrencies is regularly updated at \url{https://www.crypto51.app/}.}

Our methods for identifying these attacks initiated from our observations of the cryptocurrency space over the years, especially through Twitter.
We explored blockchain and cryptocurrency news portals and discussion fora, conducted general internet search, and searches on Twitter, to finalise the timeline through manual scrutiny.
Unlike~\cite{SSVK2019}, we have provided manually checked references for all events -- at least two cross-checked references in most cases.
With the timeline at hand, we identified a methodology to trace the events on Twitter.
We chose a common period of investigation for the events.
This started from the day before the attack to six days after the attack ended.
Thereafter, we conducted manual scrutiny to identify the keywords for searching the relevant tweets, and to clean the datasets.
These typically included the full name of the cryptocurrency, their acronyms, and keywords related to the $51\%$ attack.
For cryptocurrencies like Verge with names having general meaning, we had to use additional filtering using keywords related to cryptocurrencies.

For an attack event $E_i$ on our timeline, we have compared the volume (i.e.~number) of all tweets on the cryptocurrency ($\lvert \DSW_i \rvert$) with the volume of the subset of tweets that are explicitly about the attack ($\lvert \DSA_i \rvert$).
We have also compared these two volumes with that of a benchmark dataset $(\lvert \DSB_i \rvert)$ from a time when the respective cryptocurrency was not under attack.
In most cases, the volume has increased during the attack in comparison to the period when there is no attack.
At the time of the attack, most of the tweets are related to the attack, for every event.
This reaffirms our understanding that \textit{those interested in a cryptocurrency are certainly vigilant about its security against the $51\%$ attack as well}.

To understand the reaction times, we considered the peak day of discussion in terms of volume of tweets, counting from the day the attack ended.
We found that the \textit{delay in achieving the peak varies significantly between the events}.
For cryptocurrencies that have been attacked multiple times, the reaction time is longer for the first event.
\textit{For later events, the users are more alert and hence the peak days are closer to the day the attack ended.}

To study the sentiments of users around the events in the timeline, we have defined the \textit{sentiment profile} for a dataset to capture the percentage distribution of positive, neutral and negative sentiments as categorised by the composite score of VADER (Valence Aware Dictionary for sEntiment Reasoning) lexicon~\cite{hutto14}.
% VADER lexicon.
When the cryptocurrencies are not under attack, we found some general variation in the sentiment profile of the datasets.
While some cryptocurrencies like Bitcoin, Litecoin Cash and Vertcoin have a high degree of positive and neutral sentiments when not under attack, that can not be generalised for all cryptocurrencies on our timeline.
\textit{When under attack, however, almost every cryptocurrency succumbs to negative emotions.}

For more fine-grained analysis, we have defined the \textit{emotion profile} for a dataset.
Each tweet is individually profiled with normalised intensities of five emotions - happy, angry, surprise, sad and fear, as characterised by the Text2Emotion lexicon~\cite{Text2Emotion}.
This essentially distributes the tweets into $2^5 = 32$ sets characterised by the presence or absence of the emotions.
The emotion profile of a dataset has been defined in terms of the volume and averages of the intensities of the emotions and their combinations.
The comparison between the emotion profiles of the three types of datasets for an event as well as across events provides some general trends and deeper insights for some specific cryptocurrencies.
In particular, we note that \textit{for all cryptocurrencies, fear is a perpetually dominant emotion that intensifies at the time of an attack.}

% We have studied the sentiment and emotion profile variations between times when a cryptocurrency is under attack and at other normal (benchmark) times, between multiple attacks on the same cryptocurrency, and between different cryptocurrencies.
Our other findings are as follows.
% We confirm the common perception \textit{that those interested in a cryptocurrency are certainly vigilant about its security against the 51\% attack as well}.
% The increase in volume denotes the heightened interest of users at the time of the attack.
% Moreover, most of these tweets are related to the attack, for every attack event, justifying our conclusion.
We found common patterns across cryptocurrencies through intra-currency comparisons.
First, we found a common ``attack period'' of $6$ days for all events when there is a high volume of related tweets.
When a cryptocurrency is attacked more than once, the peak day in terms of the volume of tweets is quicker for subsequent attacks (Figure~\ref{fig:peak-day-delays}).
%We found that the sentiments are predominantly negative at the time of an attack, compared to normal periods.
We have presented inter-currency comparisons through pictorial views.
Figures~\ref{fig:sentiment-profiles-of-DSW},~\ref{fig:sentiment-profiles-of-DSA} and~\ref{fig:sentiment-profiles-of-DSB} compare the sentiment profiles of the datasets;
% The fear emotion is always predominant for all cryptocurrencies.
Figures~\ref{fig:emotion-volume-DSW},~\ref{fig:emotion-volume-DSA} and~\ref{fig:emotion-volume-DSB} compare their emotion profiles;
Figures~\ref{fig:intensities-happiness},~\ref{fig:intensities-anger},~\ref{fig:intensities-surprise},~\ref{fig:intensities-sadness} and~\ref{fig:intensities-fear} provide inter-currency comparison of the intensities of the $5$ emotions considered in the emotion profiles -- happiness, anger, surprise, sadness and fear.
We have pointed out some finer yet noteworthy points that come out of these comparisons in Sections~\ref{sec:results-sentiments},~\ref{sec:results-emotions} and~\ref{sec:results-peakdays}.

We provide detailed justifications for our methodologies in Section~\ref{sec:justification-of-methodology}.
% We have made our cleaned datasets available at~\cite{our-data}.
Our implementation is available at~\cite{our-code}.
Making our datasets public may violate X's (Twitter's) data sharing terms and policies.
They can be provided upon request.

\subsection{Importance\label{sec:importance}}
We demonstrate the frequent occurrence of $51\%$ attacks with larger cryptocurrencies being particularly targeted multiple times, thus informing against the common perception that such attacks are rather rare.
Various other attacks on cryptocurrency related systems are already known to be very common, causing huge losses to users~\cite{CPNX2020,ZQTLG21,WZYLW2022,ZXECWWQWSG2023-IEEESnP,HSB24}.

Given the ubiquity of various kinds of attacks on cryptocurrencies, a common mechanism for their detection and notification to users is of importance.
In fact, users have not defended themselves effectively due to the lack of such notifications~\cite{WZYLW2022}.
Our techniques could be used to deploy very light-weight triggering systems for detecting attacks on cryptocurrencies using social media. %, perhaps as a simple user-controlled browser extension.
Here, we clarify that the best conceivable way to detect a $51\%$ attack (or a threat thereof) early, is to keep monitoring the underlying blockchain itself for its forks and whether they are due to a $51\%$ attack~\cite{YLCGF2018}.
However, constructing such detection systems for each of the thousands of cryptocurrencies that exist today would be very costly.
Our work provides a low-cost one-point solution to this early detection problem across all cryptocurrencies.
Our technique relies on the attacks being reported first in posts by experts and enthusiasts with strong interests in the respective cryptocurrencies.
As elaborated in Section~\ref{sec:data-collection}, the use of social media for disseminating such information is common practice, justifying the feasibility of our approach for an effective solution to the problem.
More generally, for any phenomenon on a social media platform with insufficient data (but high impact like in the case of cryptocurrencies that have a trillion-dollar market), their volumetric, sentiment and emotion profiling can provide very simple yet informative analysis of key considerations like reaction times of users.

Detection of attacks and the profiling of user sentiments and emotions is important for risk analysis of cryptocurrencies.
Services like the CertiK Skynet~\cite{certik-skynet} use undisclosed and perhaps heuristic techniques (that are hard to explain) to measure user perceptions and report the ``community trust'' scores for cryptocurrencies.
Our lexicon-based approach for profiling social media data could be used to develop an aggregated score to work as an open counterpart to their score.
We hope to have set the stage for such open techniques for risk analysis that are easily understood, subject to public scrutiny and critique, and hence more explainable and trustworthy.

Finally, many governments around the world have called for regulations on cryptocurrencies\footnote{UK's promise: \url{https://www.bbc.com/news/technology-64468617}}.
Risk analysis of cryptocurrencies using open techniques have the potential to enable and/or complement such regulatory frameworks by informing the users about the risks in real-time while they are invested in cryptocurrencies.
This work is a concrete step in that direction.

%\begin{itemize}
%    \item Comparison with the first major event \blue{point out the implications on the changing sentiments on social media about the attacks}
%    \item Comparison with benchmarks \blue{point out the implications on the use of sentiment analysis as a triggering system}
%\end{itemize}

\section{Background\label{sec:brief-background}}

% The necessary backgrounds on blockchains, sentiment analysis and emotion detection have been provided in Appendix~\ref{sec:background-andprelims}.
The necessary background on blockchains has been provided briefly in Appendix~\ref{sec:background-andprelims}.
Our contributions and the results should be fairly clear without these details.
A more detailed exposition on blockchains related to this work may be found in the books~\cite{NBFMG16,CS2020-algo-finance-book}.

\subsection{Sentiment analysis background\label{sec:sentiment-analysis-background}}

% \textcolor{red}{Reviewer 2, minor issues: "2.2 and 2.3, which are supposed to be the more relevant background to the body of the work, are less technical and could benefit from bringing in the details associated with those techniques from the approach (those aren’t necessarily new techniques of the authors, and are rather a background knowledge)."}
Sentiment analysis or opinion mining is a widely used NLP technique with the aim to classify texts into typically three polarity levels, namely negative, neutral, and positive.
Such analysis can be based on a machine learning algorithm, which requires a fair amount of labelled training data to be able to classify the rest of the text into sentiment categories.
In the absence of labelled training data from the context under investigation, a rule-based lexicon (a dictionary of labelled words and phrases) is often used where words and phrases are already scored against a pre-defined set of associated sentiments by intensity (e.g., VADER or TextBlob).
As a result, such lexicon-based techniques are easily explainable.
We employed the VADER (Valence Aware Dictionary for sEntiment Reasoning) lexicon~\cite{hutto14} which has been used by many studies (such as~\cite{bonta2019compareVADER,BB2020,MUM2020,PK2020,tumasjan2021twitter,li2020incorporating}) with high accuracy on micro-blog texts such as Twitter. 
Given a text, the VADER sentiment analyser will output a compound score, aggregated from each word's valence in the text.
This score varies between $-1$ indicating that the text is extremely negative and $1$ indicating that the text is extremely positive.
Along with sentiments, the volumes of social media posts are often considered simultaneously.
For example,~\cite{oliveira2017impact} found that both sentiments and volumes are important factors for stock market price prediction.
In contrast, \cite{abraham2018cryptocurrency} found that for Bitcoin and Ethereum, the tweet volume is a better price predictor than the tweet sentiments, which in their dataset was overwhelmingly positive.
We examine both the sentiments and the volumes of tweets as indicators of user perception and reaction.
% to determine reaction day patterns and to find out how public confidence changes around these $51\%$ attacks.

\subsection{Emotion detection background\label{sec:emotion-detection-background}}

Emotion detection is an alternative way to gain insights about people’s opinions on topics.
It identifies human emotions such as anger, fear, joy, love, sadness, and surprise from texts.
Emotions on social media are known to have significant impact on people's behaviour, especially in making decisions in financial markets~\cite{ge2020beyond,griffith2020emotions}.
There are some predefined libraries for emotion detection, such as the Text2Emotion lexicon~\cite{Text2Emotion} or the NRC Emotion lexicon~\cite{NRCLex}.
As explained above, the outputs of these lexicon-based libraries are easily explainable.
The Text2Emotion lexicon outputs five types of emotions whereas the NRC Emotion lexicon outputs eight.
However, the former has the advantage of relatively simple setup and easy interpretation and has been used in different contexts with social media data \cite{bhooshan2022sentiment,chakraborty2023analysis}.
For example,~\cite{dhar2020emotions} used it to detect emotions in tweets of the top $150$ companies listed in the New York Stock Exchange - specifically, its effects on the price movements of the these companies from the Fortune $1000$ list.
They found that emotions were ``significant predictors'' for stock price movements of firms.
Similarly,~\cite{aslam2022sentiment} used the Text2Emotion lexicon to analyse general cryptocurrency related tweets and found that most tweets are associated with happy emotions, followed by fear and surprise.
The Text2Emotion Python library classifies the text into five emotions happy, angry, surprise, sad and fear, by intensity and their sum is either $0$ or $1$.
In this study we also use the Text2Emotion lexicon library to explore patterns in the emotions of tweets. % related to $51\%$ attacks on cryptocurrencies.

\begin{comment}
\section{Further Sentiment Analysis}

\begin{figure}[!htb]
\centerline{\includegraphics[width=9cm]{Figures/subset_positiveBigram_bitcoin_15_06_2014.png}}
\caption{Bigram of the positive sentiments in $\DSA_4$ subset for the event $E_4$.}
\label{fig:bigram-positive-sentiments-DSA}
\end{figure}

To help identify the reasons behind the positive sentiments on the Bitcoin [12-13 June 2014] event, we created a Bigram of the positive sentiments in $\DSA_4$ as shown in Figure~\ref{fig:bigram-positive-sentiments-DSA}.
In Natural Language Processing, Bigrams are a special case of $N$-grams (with $N=2$), showing a network or graph of word pairs that are most frequently used together in any given text.
Each word is denoted by a node in the graph and an edge between two of them shows that the pair has been used more than a threshold number of times.
At the centre of the figure the most frequently used word(s) could be detected, while closer to the sides are the less frequent ones.
Figure~\ref{fig:bigram-positive-sentiments-DSA} shows that the statements from the mining pool Ghash.io that they would ``never launch an attack'' and that ``mining pool says trust technica'' are responsible for a fair share of these positive sentiments.
\end{comment}

\section{Methodology\label{sec:methodology}}

\subsection{Data collection\label{sec:data-collection}}
Our data collection followed a two-step process.
We first conducted manual aggregation of $51\%$ attack events on various cryptocurrencies and created a timeline of these events.
Thereafter, we collected Twitter data around the identified time periods for each of those events.

\subsubsection{Creating the Timeline\label{sec:creating-timeline}}

There is no documented compilation of all $51\%$ attacks on cryptocurrencies, to the best of our knowledge.
So we have manually compiled a set of $51\%$ attack events.
We call this chronologically ordered set the \textit{timeline of events}.
Each event has a \textit{period} $(t^{s}, t^{e})$, where $t^{s}$ denotes its \textit{start date} and $t^{e}$ denotes its \textit{end date}.
% For two events $E_i, E_j \in T$, $i<j$ denotes that $E_i$ started before $E_j$ $(t^{s}_i < t^{s}_j)$, thus determining the chronology.
The chronology is determined by the start dates $t^{s}$ of the events.
We have discovered $31$ events in total.
The \textit{full timeline} is presented in Table~\ref{tab:attack-timeline-all}.
% This list may not be exhaustive, as argued at the end of this section.

There is no established methodology of creating such a timeline for cryptocurrency events.
We followed a two-pronged approach for identifying the events and their dates - spotting online artefacts, and corroborating them with social media data.
% Our methodology for constructing the timeline was the following.
We came to know of many $51\%$ attacks over the years.
Additionally, we performed comprehensive secondary research via blockchain news portals and discussion fora followed by general internet search to identify more events.
For the events thus identified, we noted the dates when they occurred.
This gave us the initial timeline of events.
We then conducted experimental searches on Twitter to find tweets related to those events around the noted dates.
In particular, we searched with the name and the acronym of a cryptocurrency along with the phrase ``51 attack''.
We manually went through the tweets to check if they indeed corroborated with the attack.
Table~\ref{tab:attack-timeline-all} contains references for all events in the full timeline.

Our methodology for constructing the timeline has its limitations.
Given the thousands of cryptocurrencies\footnote{A list is also available at~\url{https://coinmarketcap.com/all/views/all/}.} that exist today~\cite{statista2022}, we may have missed low profile $51\%$ attacks that were not captured in our searches.
In fact, attacks on less popular cryptocurrencies may not have been reported at all and hence would not be part of our results.
% That said, we also do not see a better way to detect such events.

While creating the full timeline, we observed the following.
The first known $51\%$ attack on a cryptocurrency was on Feathercoin in $2013$~\cite{AttackFeathercoin2013-1,AttackFeathercoin2013-2}.
The first major event concerning a possible $51\%$ attack on a cryptocurrency that grabbed popular attention was in $2014$ for Bitcoin~\cite{AttackBitcoin2014-1,AttackBitcoin2014-2,AttackBitcoin2014-3}.
% Bitcoin, the most popular cryptocurrency of all times faced such a threat in $2014$, when one of the miners attained around $50\%$ of the computational power, raising major alarms in the cryptocurrency space~\cite{AttackBitcoin2014-1,AttackBitcoin2014-2,AttackBitcoin2014-3}.
After the huge jump in the prices of major cryptocurrencies like Bitcoin in $2017$, there were a wide variety of high profile and low profile $51\%$ attacks. % (in  terms of the market capitalisation of the cryptocurrencies that were attacked, the number of news articles published on those attacks, and the amount of discussions that followed these attacks on social media).
Twitter became a common medium for real-time dissemination of news and opinions on cryptocurrency events (much like for other financial instruments~\cite{yang2015twitter,afanasyev2021strength,gross2019buzzwords,behrendt2018twitter,mohan2019stock}) since around that time.
Influential Twitter profiles such as that of Vitalik Buterin, one of the creators of Ethereum, started tweeting on $51\%$ attacks and other events. %, while other attacks went almost unrecognised.
Hence we chose Twitter data for our analysis.

\subsubsection{Gathering Twitter Data\label{sec:gathering-twitter-data}}

For our comparative analysis, we selected $17$ of the $31$ events, labelled with $E_i$ in Table~\ref{tab:attack-timeline-all}.
We call this shortened list as the \textit{reduced timeline} $T$.
These are events of \textit{cryptocurrencies that were attacked more than once}, so that comparisons between events of the same cryptocurrency is possible.
The only exceptions are the events of Feathercoin in $2013$~\cite{AttackFeathercoin2013-1,AttackFeathercoin2013-2} and Bitcoin in $2014$~\cite{AttackBitcoin2014-1,AttackBitcoin2014-2,AttackBitcoin2014-3}, that have been included for their significance explained in Section~\ref{sec:creating-timeline}.

We have gathered Twitter data for events in the reduced timeline $T$.
We chose an \textit{attack data period} $(t^s-1, t^e+6)$ for collecting the data on each $E_i \in T$.
This meant that we searched for tweets from \textit{one day before}, until \textit{one week after} the event period.
For example, the period of the event $E_4$ was from 16 May 2018 to 19 May 2018 when Bitcoin Gold was attacked.
So we collected tweets for this event for the duration 15 May 2018 (i.e.~one day before the start date 16 May 2018 of $E_4$) until 25 May 2018 (i.e.~one week after the end date 19 May 2018 of $E_4$).

\begin{figure}[!htb]
\centerline{\includegraphics[width=9cm]{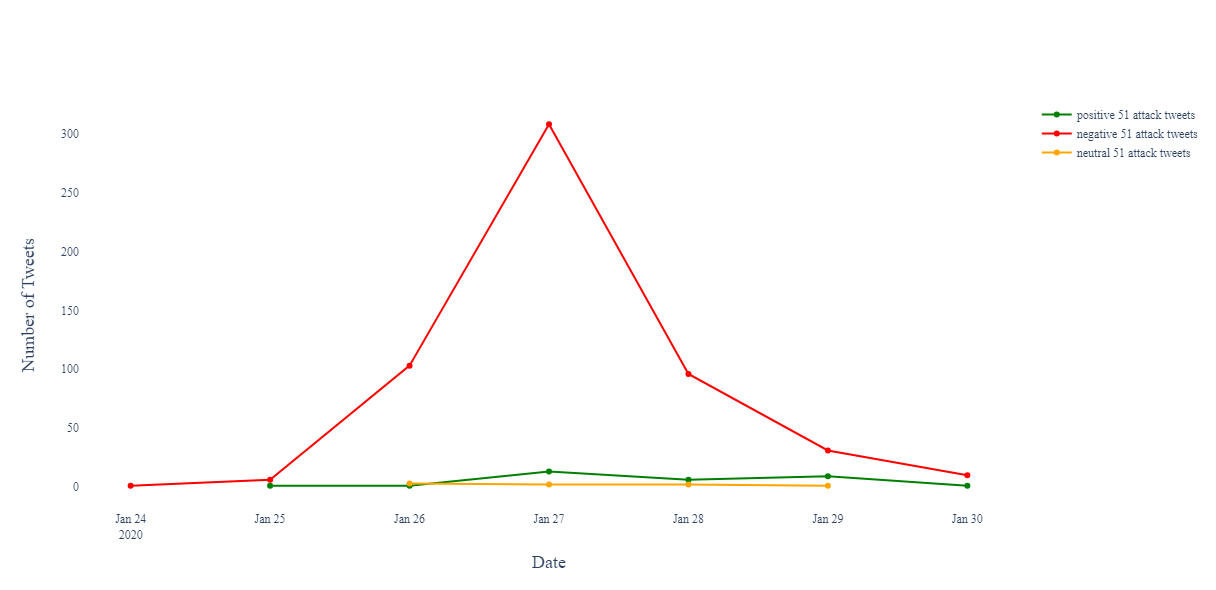}}
\caption{
Timeline of sentiments in $\DSA_{12}$ for event $E_{12}$ of the second attack on Bitcoin Gold [23-24 January 2020].
The plot of the sentiment-wise volumes of tweets show that the peak was achieved on 27 January with over $300$ negative tweets.
Therefore, the peak day of this particular event is on the third day following the end date of the $51\%$ attack.
}
\label{fig:peak-day-E12}
\end{figure}

We started from $t^s-1$ to provide a buffer to the start date $t^s$ as reported in the references we collected for each event.
Our choice for $t^e+6$ is based on our observation on the volume of tweets of all events $E_i$.
As an example, consider Figure~\ref{fig:peak-day-E12} for the attack event $E_{12}$ on Bitcoin Gold.
It shows that the volume of tweets keeps growing until it reaches a peak and then goes back to almost the pre-attack levels within $6$ days from the day the attack ended.
This is interestingly true for almost all events, thus providing a rationale for the common period.
Also note from Figure~\ref{fig:peak-day-E12} that most tweets are negative during this period.

We used the full name of the cryptocurrency (e.g.~``Bitcoin Gold'' for $E_{12}$) as the keyword to search for relevant tweets  using the Twitter API~\cite{TwitterAPIv2}.
For each event $E_i \in T$, we created three datasets as follows.
\begin{itemize}
    \item \textit{A whole dataset} $\DSW_i$ for the attack data period $(t^s_i - 1, t^e_i + 6)$,
    \item \textit{An attack dataset} $\DSA_i$ that is a subset of the whole dataset $\DSW_i$, containing tweets \textit{explicitly discussing the attack}, for the attack data period $(t^s_i - 1, t^e_i + 6)$, and
    \item \textit{A benchmark dataset} $\DSB_i$ containing all tweets for a \textit{benchmark data period}
    % $(t^s_i - 31, t^e_i - 23)$ 
    that is a \textit{month before the period of the event}.
\end{itemize}
A $51\%$ attack is so fundamental to a PoW cryptocurrency, that an attack on one PoW cryptocurrency may affect user reactions on other PoW cryptocurrencies as well.
So we had to be careful that the benchmark data periods did not coincide with any of the attack data periods.
Hence, for the events $E_{15}$ of Ethereum Classic
% benchmark period [28 Sep - 4 Oct 2020]
and $E_{17}$ of Bitcoin SV,
% benchmark period [2-9 Sep 2021],
the benchmark data were taken a month after the attacks, as the benchmark data periods would otherwise coincide with some other attack data periods.

\begin{comment}
\textcolor{red}{From the gathered triples of datasets $(\DSW_i, \DSA_i, \DSB_i), 1 \leq i \leq 22$ for each event, two had insufficient data points to carry out the analyses.}
\begin{itemize}
    \item \textcolor{red}{Powercoin, which suffered a $51\%$ attack between 15-16 June 2013, had only $9$ tweets.}
    \item \textcolor{red}{Expanse, which was attacked on 29 July 2019 had only $13$ tweets.}
\end{itemize}
\textcolor{red}{Therefore these two events were left out from the timeline $T$ and this study.
The remaining events constitute our revised timeline $T$ as shown in Figure~\ref{fig:timeline-dataset-sizes}.}
\end{comment}

\subsubsection{Data Preparation}\label{sec:data-preparation}
After creating the \textit{initial raw datasets}, we conducted manual inspection on samples of collected tweets to check their suitability for our studies.
We found that the data required some cleaning and in some cases we needed to refine our search to get more appropriate data.
For cryptocurrencies like Bitcoin whose names are unambiguous, there was no need to use extra keywords to remove tweets that were off-topic.
However, in some cases such as ``Verge'' or ``Expanse'', where the name of the cryptocurrency is a general word with other common meanings, additional keywords were used to refine the search and only keep the relevant tweets in the results.
These additional keywords were the coin abbreviations (such as XVG and EXP), as well as words related to cryptocurrencies and their mining - ``crypto'', ``coin'', ``currency'', ``miner'', and ``mining''.
Only those tweets that included at least one of these terms were kept for further analysis.
This strategy was used for all three types of datasets - $\DSW_i$, $\DSA_i$ and $\DSB_i$ - for an event $E_i \in T$.
The attack datasets $\DSA_i$ were created as a subset of the whole dataset $\DSW_i$ only containing tweets that contained reference to $51\%$ attack that is also called the \textit{double spending attack}.
So these were tweets containing at least one of the keywords ``51'', ``attack'', ``double spend'' or ``double spending''.

We also cleaned the datasets to leave out non-English tweets and duplicates from the same Twitter handles.
We then applied the typical text mining based cleaning steps: removal of (1) url links, (2) special characters, and (3) extra white spaces.
We finally converted all alphabets to the lower case.

\subsection{Sentiment profiling\label{subsec:sentiment}}

We have used the VADER lexicon's compound scores to classify tweets into positive, neutral and negative sentiments~\cite{hutto14}.
% based on the suggested threshold of $0.05$ by the developers~\cite{hutto14}.
Building upon the classification of individual tweets through their compound scores, we define the following to characterise a dataset.

\begin{definition}
\label{def:sentiment-profile-dataset}
Let $\DS = (u_1, \ldots, u_n)$ be a text dataset with $n$ units of text data $u_i$, each with a compound score of $-1 \leq s_i \leq 1$.
Let $\delta_p$ denote the threshold above which the score is considered positive and $\delta_n$ is the threshold below which the score is considered negative.
The \textit{sentiment profile} of $\DS$ is defined as \[\SP(\DS) = (N(\DS), Z(\DS), P(\DS)),\]
where
\begin{align*}
    P(\DS) & = \left\lvert \{u_i : s_i \geq \delta_p, u_i \in \DS\} \right\rvert \times (100/n), \\
    Z(\DS)  & = \left\lvert \{u_i : s_i > \delta_n, s_i < \delta_p, u_i \in \DS\} \right\rvert \times (100/n), \text{ and} \\
    N(\DS) & = \left\lvert \{u_i : s_i \leq \delta_n, u_i \in \DS\} \right\rvert \times (100/n).
\end{align*}
\end{definition}

Here, $N(\DS), Z(\DS)$ and $P(\DS)$ denote the percentages of $u_i$'s in $\DS$ that are negative, neutral and positive respectively.
Note that $0 \leq N(\DS), Z(\DS), P(\DS) \leq 100$ and $N(\DS) + Z(\DS) + P(\DS) = 100$.
% For our analysis, $\delta_n = -0.05$ and $\delta_p = 0.05$.
For our analysis, $\delta_n = 0$ and $\delta_p = 0$.
Different positive and negative thresholds could be used to tune the system depending on its context.

\begin{comment}
We analysed and compared datasets through their respective sentiment profiles as defined above.
Given the nature of the dataset, certain keywords may have had a strong influence on the compound score of a tweet.
For example, an extremely negative tweet from the Bitcoin Gold 2018 event $E_4$ says, \textit{``bitcoin gold hit by double spend attack  exchanges lose millions  attack  bitcoin  gold  double  spend  attack''}.
It received a compound score of $-0.9$.
The word ``attack'' has a high intensity negative score of $-0.48$ by itself.
However, even after the three occurrences of the word ``attack'' were removed from the tweet, the total compound score still remained significantly negative at $-0.4$.
For robustness check, we have performed a comparison of compound scores on this same $51\%$ attack subset $\DSA_4$ (Bitcoin Gold attack between 16-19 May 2018) with and without the word ``attack''.
We found that even though the overall distribution of sentiments are more moderated without the word, the negative sentiments are still dominating.
Hence we have left the word ``attack'' in for the analyses of the subsequent datasets.
% Figures \ref{fig20} and \ref{fig21} can be found in the Appendix.
\end{comment}

\subsection{Emotion profiling\label{subsec:emotion}}

The Text2Emotion lexicon scores five emotions - happy, angry, surprise, sad and fear - between $0$ and $1$ and their sum is $0$ when the intensity of each emotion is $0$, and $1$ otherwise.
We are not only interested in the dominant emotion in a dataset, but the variation of all five emotions in the tweets as well (as in~\cite{aslam2022sentiment}).
We have profiled the emotions by their \textit{volume} and \textit{intensity}.

For any real number $x \in \mathbb{R}$, \textit{ceiling of} $x$ denoted as $\lceil x \rceil$ is the smallest integer greater than or equal to $x$.
\begin{comment}
In particular, for $0 \leq x \leq 1$,
\begin{align*}
    \lceil x \rceil =
    \begin{cases}
        0, \text{ if } x = 0; \\
        1, \text{ otherwise}.
    \end{cases}
\end{align*}
\end{comment}
For a data unit $u_i \in \DS$, let $h_i$ denote the intensity of the happy emotion, $a_i$ for anger, $s_i$ for surprise, $d_i$ for sadness, and $f_i$ for fear.
\begin{definition}
\label{def:emotion-profile-dataset}
Let $\DS = (u_1, \ldots, u_n)$ be a text dataset with $n$ units of text data.
The \textit{emotion intensity} of a data unit $u_i$ is defined as
\[\ei(u_i) = (h_i, a_i, s_i, d_i, f_i)\]
such that $0 \leq h_i, a_i, s_i, d_i, f_i \leq 1$ and
\begin{equation*}
h_i + a_i + s_i + d_i + f_i =
\begin{cases}
0, \text{ if } h_i = a_i = s_i = d_i = f_i = 0; \\
1, \text{ otherwise}.
\end{cases}
\end{equation*}
The emotion intensity of the dataset $\DS$ is defined as
\[\EI(\DS) = (H(\DS), A(\DS), S(\DS), D(\DS), F(\DS))\]
such that
\begin{align*}
  & H(\DS) = \frac{1}{n}\sum_{i=1}^{n} h_i,
    A(\DS) = \frac{1}{n}\sum_{i=1}^{n} a_i,
    S(\DS) = \frac{1}{n}\sum_{i=1}^{n} s_i, \\
  & D(\DS) = \frac{1}{n}\sum_{i=1}^{n} d_i, \text{ and }
    F(\DS) = \frac{1}{n}\sum_{i=1}^{n} f_i.
\end{align*}
The emotion volume of a data unit $u_i$ is defined as
\[\ev(u_i) = (hv_i, av_i, sv_i, dv_i, fv_i)\]
such that 
\begin{equation*}
hv_i = \lceil h_i \rceil, av_i = \lceil a_i \rceil, sv_i = \lceil s_i \rceil, dv_i = \lceil d_i \rceil, fv_i = \lceil f_i \rceil.
\end{equation*}
The emotion volume of the dataset $\DS$ is defined as
\[\EV(\DS) = (HV(\DS), AV(\DS), SV(\DS), DV(\DS), FV(\DS))\]
such that
\begin{align*}
  & HV(\DS) = \sum_{i=1}^{n} hv_i,
    AV(\DS) = \sum_{i=1}^{n} av_i,
    SV(\DS) = \sum_{i=1}^{n} sv_i, \\
  & DV(\DS) = \sum_{i=1}^{n} dv_i, \text{ and }
    FV(\DS) = \sum_{i=1}^{n} fv_i.
\end{align*}
The emotion profile of a dataset $\DS$ is defined as
\[\EP(\DS) = (\EI(\DS), \EV(\DS)).\]
\end{definition}

For a data unit (i.e.~a tweet) $u_i$, even if only one of $h_i, a_i, s_i, d_i, f_i$ has a non-zero value, then $h_i + a_i + s_i + d_i + f_i = 1$.
If there are two non-zero emotions, each carrying the same intensity, their values would be $0.5$ each, and so on.
% Similarly, if there are five non-zero emotions, each of the same intensity, their values would be $0.2$ each.
\begin{comment}
For example, in the cleaned whole dataset $E_4$ of Bitcoin Gold [16-19 May 2018], out of $2449$ tweets there were $661$ tweets with non-zero emotions of equal intensity.
That is nearly $27\%$ of the data points in the dataset.
\end{comment}
For a dataset $\DS$, $\EI(\DS)$ denotes its emotion intensity.
Each element in the tuple $\EI(\DS)$ is an average of the intensities of a corresponding emotion in all tweets in the dataset:
$H(\DS)$ is the average for happy, $A(\DS)$ is for anger, $S(\DS)$ is for surprise, $D(\DS)$ is for sadness, and $F(\DS)$ is for fear.
A dataset gets characterised by these mean values in its emotion intensity.
The emotion volume $\EV(\DS)$ of the dataset has counts of the number of data units $u_i \in \DS$ containing the five emotions:
$HV(\DS)$ is the number for happy, $AV(\DS)$ is for angry, $SV(\DS)$ is for surprise,  $DV(\DS)$ is for sadness, and $FV(\DS)$ is for fear.
The emotion profile $\EP(\DS)$ of the dataset is the pair $(\EI(\DS), \EV(\DS))$.
We compare the events $E_i \in T$ using the emotion profile of their datasets $\DSW_i, \DSB_i$ and $\DSA_i$.

\subsection{Justification of Methodology\label{sec:justification-of-methodology}}
We have created the timeline presented in Table~\ref{tab:attack-timeline-all} manually, with more than one cross-checked references in most cases.
Note that the only other previous work~\cite{SSVK2019} providing such a (albeit much smaller) list did not provide such references, nor did they mention how they created their list.
This perhaps reflects the difficulty in identifying appropriate sources that could be used to create such a list.
We do not claim our manual search to be fool-proof or our list to be exhaustive.
However, we also do not see an easy way to automate the creation of such a list with limited resources, that would be exhaustive.
Our sources can be found from our references in Table~\ref{tab:attack-timeline-all}.
Being a first attempt to create such a timeline since the introduction of Bitcoin in 2009, our method demonstrates that such a timeline can indeed be created and would reveal the vulnerability of the cryptocurrency ecosystem in the face of $51\%$ (and other) attacks.

Our data collection methodology includes a preliminary step for eliminating off-topic or irrelevant tweets from all our datasets.
We analysed the time frame of each attack manually and found a common ``attack period'' --  the time interval of data collection.
Figure~\ref{fig:peak-day-E12} for the attack event $E_{12}$ on Bitcoin Gold is an example demonstrating the rise and fall in the volume of tweets, justifying our choice.
We have also conducted significant manual checks to ensure that our methods do not usually generate false positive or false negative results.
For example, we have manually checked the dataset $\DSW_4$ with and without the word ``attack'' and found that although the overall compound scores are more moderate without the word ``attack'', the dominant sentiment is still negative.
The reason for comparing the whole datasets $\DSW_i$ with their corresponding attack datasets $\DSA_i$ was precisely to check the influence of keywords like ``attack'' and ``double spending'' on the sentiments and emotions detected by these systems.
We would also emphasise that an attack event is detrimental to the trust that people (supporters and detractors alike) have on the security of the blockchain.
So the majority of negative sentiments corroborate with our expectations from the analysis. 

Finally, we justify our choice of the lexicon-based approach in place of heavier (machine learning based) techniques.
The tools (VADER and Text2Emotion) we have chosen are very simple, but very effective for our goals, as demonstrated by the consistent behaviour of our findings across cryptocurrencies.
We were interested to see if an attack can be detected at all from social media, and if so, what patterns emerge from them.
In our analysis we worked with $51$ datasets (many are quite small with only tens or hundreds of data units), which gave a rationale to use these rule-based approaches.
Our conscious choice of the tools are based on the following reasons.
\begin{itemize}
    \item Transparency:
    Simple lexicon-based packages such as VADER and Text2Emotion offer transparency for our metrics.
    They allow easy reproduction of our results.
    \item Works on small datasets:
    They can be easily used for attacks on smaller cryptocurrencies that do not generate enough data for more sophisticated statistical analysis systems.
    (Tables~\ref{tab:dataset-size-attack} and~\ref{tab:dataset-size-benchmark} show that events $E_1, E_6, E_9, E_{10}$ and $E_{16}$ have data sets of size less than $100$.)
    \item High accuracy: We have manually checked 100 random samples of tweets from the Bitcoin Gold dataset $\DSW_4$ and scored the validity of the chosen libraries.
    We scored a tweet $1$ if the sentiment or emotion was justified, and $0$ otherwise.
    We found that both VADER and Text2Emotion libraries perform with $81\%$ accuracy\footnote{Our manual check may be verified from the file ``accuracy\_check\_sample\_BTG\_2018.xlsx'' in our data repository~\cite{our-data}.}.
    \item Previously,~\cite{aslam2022sentiment} used the Text2Emotion library for annotation (labelling) social media data on cryptocurrencies. % to conduct general sentiment analysis and emotion detection.
    We emphasise that they have used the Text2Emotion library considering its outputs as ground truth for them to train the neural network models later.
    They have reported higher accuracy of this library than the corresponding machine learning model.
\end{itemize}
Analysing the attack events with more sophisticated statistical tools is likely to yield deeper insights.
However, as a first work establishing methodologies for investigating the attack events, perhaps more granular results from such tools would have been difficult to interpret and analyse.
In particular, we have found consistent patterns across not just the largest PoW cryptocurrencies and different periods of time, but for the diverse set of cryptocurrencies included in our analysis, hence demonstrating good generalisability of our conclusions.
Clear patterns emerged from our analysis showing common ``attack periods'', peak days, predominantly negative sentiments at the time of attack, and the fear emotion prevailing all the time for all cryptocurrencies.

\section{The Results\label{sec:experimental-results}}

\subsection{Timeline of 51\% attacks\label{sec:result-timeline}} 

The $31$ attacks we discovered are arranged chronologically in Table~\ref{tab:attack-timeline-all}.
These are attacks on $20$ different cryptocurrencies, as listed in Table~\ref{tab:cryptocurrencies-attacked-number-of-times}. % of Appendix~\ref{sec:further-details-on-timeline}.
The events for the top $8$ cryptocurrencies in Table~\ref{tab:cryptocurrencies-attacked-number-of-times} have been included in our analysis of the reduced timeline $T$.
The first $6$ of them have been attacked more than once, allowing us to compare between events of the same cryptocurrency.
The attacks on Feathercoin and Bitcoin have been included because of their special significance as explained in Section~\ref{sec:creating-timeline}.
Hence our analysis is on $\#T = 17$ events and $51$ datasets - three datasets $\DSW_i, \DSA_i, \DSB_i$, for each event $E_i \in T, 1 \leq i \leq \#T$.
The event $E_4$ is when Bitcoin faced only a threat of a $51\%$ attack, but was not actually attacked.
The event $E_{12}$ is for an attack on Bitcoin Gold that was averted.
Table~\ref{tab:dataset-size-attack} presents the reduced timeline $T$ as a set of events $E_1, \ldots, E_{17}$ ordered chronologically, along with references to news articles and other relevant weblinks that corroborate their occurrences.
Table~\ref{tab:dataset-size-attack} also includes the attack data periods $(t^s-1, t^e+6)$ and the volume of data for each event $E_i \in T$.
The columns for $\lvert \DSW_i \rvert$ and $\lvert \DSA \rvert$ contain the number of tweets in the whole dataset $\DSW_i$ and the attack dataset $\DSA_i$ respectively, for each $E_i \in T$.
Similarly, Table~\ref{tab:dataset-size-benchmark} provides the number of tweets $\lvert \DSB_i \rvert$ in the benchmark datasets $\DSB_i$ along with the dates for benchmark data collection.
The volumes of tweets in the benchmark datasets $\DSB_i$ in Table~\ref{tab:dataset-size-benchmark} are indicative of the \textit{general relative popularity of the respective cryptocurrency}.
The volumes in Table~\ref{tab:dataset-size-attack} denote the more specific \textit{popularity of the attack events themselves}.
Comparing the volumes $\lvert \DSW_i \rvert, \lvert \DSA_i \rvert$ and $\lvert \DSB_i \rvert$ we see that in most cases, the number of tweets increase during the attack in comparison to the period when there is no attack.
At the time of the attack, most of the tweets are related to the attack, for every event.
This reaffirms our understanding that \textit{those interested in a cryptocurrency are certainly vigilant about its security against the $51\%$ attack as well}.
For cryptocurrencies that have been attacked or threatened multiple times, \textit{the first event had the largest volume, and it kept decreasing in subsequent events}.
This is true for benchmark datasets as well, \textit{indicating reduced enthusiasm in the respective cryptocurrencies, after the attacks}.

\begin{table}[!htb]
\scriptsize
\caption{The full timeline of all $51\%$ attack events we identified, with their attack periods $(t^s, t^e)$. \\
Events `$E_i$' are part of our analysis; others labelled with `-'. \\
$^{**}$These events were either threats of or $51\%$ attacks that were averted.}
\begin{center}
\begin{tabular}{llll}
\hline 
Sl.No. & Event & Cryptocurrency & Attack Period $(t^s, t^e)$ \\
\hline

01 & $E_{1}$  & Feathercoin~\cite{AttackFeathercoin2013-1,AttackFeathercoin2013-2}
& (08 Jun 2013, 08 Jun 2013) \\

02 & - & Powercoin~\cite{SSVK2019,attack1powercoin-1,attack1powercoin-2} & (15 Jun 2013, 16 Jun 2013) \\
% $E_{2}$  & Bitcoin
% & (25 Sep 2013, 27 Sep 2013) & $27868$ & $301$  \\

03 & - & Terracoin~\cite{SSVK2019,AttacksKomodo_noauthor_51_2018} & (25 Jul 2013, 25 Jul 2013) \\

04 & $E_{2}$  & $^{**}$Bitcoin~\cite{AttackBitcoin2014-1,AttackBitcoin2014-2,AttackBitcoin2014-3}
& (12 Jun 2014, 13 Jun 2014) \\
% & (16 Jun 2014, 16 Jun 2014) & $72458$ & $3409$ \\

05 & - & Shift~\cite{SSVK2019,attack1krypton_shift-1} & (25 Aug 2016, 25 Aug 2016) \\

06 & - & Krypton~\cite{SSVK2019,attack1krypton_shift-1,attack1krypton_2016} & (26 Aug 2016, 26 Aug 2016) \\

07 & - & Electroneum~\cite{SSVK2019,attack1electroneum} & (04 Apr 2018, 04 Apr 2018) \\

08 & $E_{3}$  & Verge~\cite{Attack1Verge2018-1,Attack1Verge2018-2,Attack1Verge2018-3,Attack1Verge2018-4}
% & (04 Apr 2018, 04 Apr 2018) & $6155$  & $599$  \\
& (04 Apr 2018, 06 Apr 2018) \\

09 & $E_{4}$  & Bitcoin Gold~\cite{Attack1BitcoinGold2018-1,Attack1BitcoinGold2018-2}
& (16 May 2018, 19 May 2018) \\

10 & - & MonaCoin~\cite{SSVK2019,attack1_monacoin_2018} & (17 May 2018, 17 May 2018) \\

11 & $E_{5}$  & Verge~\cite{Attack2Verge2018-1,Attack2Verge2018-2,Attack2Verge2018-3}
& (21 May 2018, 22 May 2018) \\

12 & $E_{6}$  & Litecoin Cash~\cite{Attack1LitecoinCash2018-1}
& (30 May 2018, 30 May 2018)  \\

13 & - & ZenCash~\cite{Attack1Zencash2018-1,Attack1Zencash2018-2,Attack1Zencash2018-3} & (02 Jun 2018, 03 Jun 2018) \\

14 & - & FLO~\cite{AttacksKomodo_noauthor_51_2018,Attack1FLO_2018} & (08 Sep 2018, 08 Sep 2018) \\

15 & - & Pigeoncoin~\cite{SSVK2019,attack1_pigeoncoin_2018} & (26 Sep 2018, 26 Sep 2018) \\

16 & - & Bitcoin Private~\cite{SSVK2019,attack1_bitcoinprivate_2018} & (19 Oct 2018, 19 Oct 2018) \\

17 & - & Karbo~\cite{SSVK2019,attack1_karbo} & (10 Nov 2018, 10 Nov 2018) \\

18 & - & AurumCoin~\cite{AttacksKomodo_noauthor_51_2018,Attack1AurumCoin-2,Attack1Aurum-3} & (11 Nov 2018, 11 Nov 2018) \\

19 & $E_{7}$  & Vertcoin~\cite{Attack1Vertcoin2018-1,Attack1Vertcoin2018-2}
% & (12 Oct 2018, 02 Dec 2018) & $1596$  & $382$  \\
& (02 Dec 2018, 02 Dec 2018) \\

20 & $E_{8}$ & Ethereum Classic~\cite{Attack1EthereumClassic2019-1,Attack1EthereumClassic2019-2}
& (05 Jan 2019, 08 Jan 2019) \\

21 & $E_{9}$ & Litecoin Cash~\cite{Attack2LitecoinCash2019-1}
& (04 Jul 2019, 07 Jul 2019) \\

22 & - & Expanse~\cite{attack1_exp} & (29 Jul 2019, 29 Jul 2019) \\

23 & $E_{10}$ & Vertcoin~\cite{Attack2Vertcoin2019-1,Attack2Vertcoin2019-2,Attack2Vertcoin2019-3}
& (01 Dec 2019, 01 Dec 2019) \\

24 & $E_{11}$ & Bitcoin Gold~\cite{Attack2BitcoinGold2020-1,Attack2BitcoinGold2020-2,Attack2BitcoinGold2020-3}
& (23 Jan 2020, 24 Jan 2020) \\

25 & $E_{12}$ & $^{**}$Bitcoin Gold~\cite{Attack3BitcoinGold2020-1,Attack3BitcoinGold2020-2}
& (02 Jul 2020, 10 Jul 2020) \\

26 & $E_{13}$ & Ethereum Classic~\cite{Attack2EthereumClassic2020-1,Attack2EthereumClassic2020-2,Attack2EthereumClassic2020-3}
& (29 Jul 2020, 01 Aug 2020) \\

27 & $E_{14}$ & Ethereum Classic~\cite{Attack3EthereumClassic2020-1,Attack3EthereumClassic2020-2}
& (05 Aug 2020, 05 Aug 2020) \\

28 & $E_{15}$ & Ethereum Classic~\cite{Attack4EthereumClassic2020-1,Attack4EthereumClassic2020-2}
& (29 Aug 2020, 29 Aug 2020) \\

29 & - & Firo~\cite{Attack1Firo2021-1,Attack1Firo2021-2,Attack1Firo2021-3} & (20 Jan 2021, 20 Jan 2021) \\

30 & $E_{16}$ & Bitcoin SV~\cite{Attack1BitcoinSV2021-1}
& (24 Jun 2021, 09 Jul 2021) \\

31 & $E_{17}$ & Bitcoin SV~\cite{Attack2BitcoinSV2021-1,Attack2BitcoinSV2021-2}
& (03 Aug 2021, 03 Aug 2021) \\      

\hline
\end{tabular}
\label{tab:attack-timeline-all}
\end{center}
\end{table}

\begin{table}[!htb]
\scriptsize
\caption{Attack data periods $(t^s-1, t^e+6)$ and dataset sizes of $\DSW_i$ and $\DSA_i$, for all $E_i \in T, 1 \leq i \leq \#T$.}
\begin{center}
\begin{tabular}{lllrr}
\hline 
Event & Cryptocurrency & Att.~Dat.~Per.~$(t^s-1, t^e+6)$ & $\lvert \DSW_i \rvert$ & $\lvert \DSA_i \rvert$ \\
\hline

$E_{1}$  & Feathercoin~\cite{AttackFeathercoin2013-1,AttackFeathercoin2013-2}
& (07 Jun 2013, 14 Jun 2013) & $261$   & $26$   \\

$E_{2}$  & $^{**}$Bitcoin~\cite{AttackBitcoin2014-1,AttackBitcoin2014-2,AttackBitcoin2014-3}
& (11 Jun 2014, 19 Jun 2014) & $72458$ & $3409$ \\

$E_{3}$  & Verge~\cite{Attack1Verge2018-1,Attack1Verge2018-2,Attack1Verge2018-3,Attack1Verge2018-4}
& (03 Apr 2018, 12 Apr 2018) & $6155$  & $599$  \\

$E_{4}$  & Bitcoin Gold~\cite{Attack1BitcoinGold2018-1,Attack1BitcoinGold2018-2}
& (15 May 2018, 25 May 2018) & $2449$  & $1012$ \\

$E_{5}$  & Verge~\cite{Attack2Verge2018-1,Attack2Verge2018-2,Attack2Verge2018-3}
& (20 May 2018, 28 May 2018) & $3661$  & $685$  \\

$E_{6}$  & Litecoin Cash~\cite{Attack1LitecoinCash2018-1}
& (29 May 2018, 05 June 2018) & $842$   & $28$   \\

$E_{7}$  & Vertcoin~\cite{Attack1Vertcoin2018-1,Attack1Vertcoin2018-2}
& (01 Dec 2018, 08 Dec 2018) & $1596$  & $382$  \\

$E_{8}$ & Ethereum Classic~\cite{Attack1EthereumClassic2019-1,Attack1EthereumClassic2019-2}
& (04 Jan 2019, 14 Jan 2019) & $6840$  & $4465$ \\

$E_{9}$ & Litecoin Cash~\cite{Attack2LitecoinCash2019-1}
& (03 Jul 2019, 13 Jul 2019) & $462$   & $58$   \\

$E_{10}$ & Vertcoin~\cite{Attack2Vertcoin2019-1,Attack2Vertcoin2019-2,Attack2Vertcoin2019-3}
& (30 Nov 2019, 07 Dec 2019) & $457$   & $378$  \\

$E_{11}$ & Bitcoin Gold~\cite{Attack2BitcoinGold2020-1,Attack2BitcoinGold2020-2,Attack2BitcoinGold2020-3}
& (22 Jan 2020, 30 Jan 2020) & $941$   & $594$  \\

$E_{12}$ & $^{**}$Bitcoin Gold~\cite{Attack3BitcoinGold2020-1,Attack3BitcoinGold2020-2}
& (01 Jul 2020, 16 Jul 2020) & $712$   & $218$  \\

$E_{13}$ & Ethereum Classic~\cite{Attack2EthereumClassic2020-1,Attack2EthereumClassic2020-2,Attack2EthereumClassic2020-3}
& (28 Jul 2020, 07 Aug 2020) & $1691$  & $900$  \\

$E_{14}$ & Ethereum Classic~\cite{Attack3EthereumClassic2020-1,Attack3EthereumClassic2020-2}
& (04 Aug 2020, 11 Aug 2020) & $1649$  & $967$  \\

$E_{15}$ & Ethereum Classic~\cite{Attack4EthereumClassic2020-1,Attack4EthereumClassic2020-2}
& (28 Aug 2020, 04 Sep 2020) & $1121$  & $706$  \\

$E_{16}$ & Bitcoin SV~\cite{Attack1BitcoinSV2021-1}
& (23 Jun 2021, 15 Jul 2021) & $2174$  & $96$   \\

$E_{17}$ & Bitcoin SV~\cite{Attack2BitcoinSV2021-1,Attack2BitcoinSV2021-2}
& (02 Aug 2021, 09 Aug 2021) & $1159$  & $524$  \\      

\hline
\end{tabular}
\label{tab:dataset-size-attack}
\end{center}
\end{table}

\begin{table}[!htb]
\scriptsize
\caption{Benchmark data periods and dataset sizes of $\DSB_i$, for all $E_i \in T, 1 \leq i \leq \#T$.}
\begin{center}
\begin{tabular}{lllr}
\hline 
Event & Cryptocurrency & Benchmark Data Period & $\lvert \DSB_i \rvert$ \\
\hline
$E_{1}$  & Feathercoin      & (07 May 2013, 16 May 2013) & $51$     \\
%$E_{2}$  & Bitcoin          & (24 Aug 2013, 03 Sep 2013) & $23725$  \\
$E_{2}$  & $^{**}$Bitcoin          & (15 May 2014, 22 May 2014) & $65408$  \\
$E_{3}$  & Verge            & (03 Mar 2018, 10 Mar 2018) & $7804$   \\
$E_{4}$  & Bitcoin Gold     & (15 Apr 2018, 25 Apr 2018) & $1137$   \\
$E_{5}$  & Verge            & (20 Apr 2018, 28 Apr 2018) & $4030$   \\
$E_{6}$  & Litecoin Cash    & (29 Apr 2018, 05 May 2018) & $1338$   \\
$E_{7}$  & Vertcoin         & (11 Sep 2018, 21 Sep 2018) & $297$    \\
$E_{8}$ & Ethereum Classic & (04 Dec 2018, 14 Dec 2018) & $1105$   \\
$E_{9}$ & Litecoin Cash    & (03 Jun 2019, 13 Jun 2019) & $542$    \\
$E_{10}$ & Vertcoin         & (30 Oct 2019, 07 Nov 2019) & $65$     \\
$E_{11}$ & Bitcoin Gold     & (22 Dec 2019, 30 Dec 2019) & $293$    \\
$E_{12}$ & $^{**}$Bitcoin Gold     & (01 Jun 2020, 16 Jun 2020) & $482$    \\
$E_{13}$ & Ethereum Classic & (28 Jun 2020, 07 Jul 2020) & $246$    \\
$E_{14}$ & Ethereum Classic & (04 Jul 2020, 11 Jul 2020) & $215$    \\
$E_{15}$ & Ethereum Classic & (28 Sep 2020, 04 Oct 2020) & $172$    \\
$E_{16}$ & Bitcoin SV       & (23 May 2021, 15 Jun 2021) & $1625$   \\
$E_{17}$ & Bitcoin SV       & (02 Sep 2021, 09 Sep 2021) & $763$    \\
\hline
\end{tabular}
\label{tab:dataset-size-benchmark}
\end{center}
\end{table}

%\begin{figure}[!htb]
%\centerline{\includegraphics[width=9cm]{Figures/DatasetSizes.png}}
%\caption{The volume of tweets in logarithmic scale for each dataset $\DSB_i$, $\DSW_i$ and $\DSA_i$ for each event $E_i \in T, 1 \leq i \leq \#T$ in the timeline $T$.}
%\label{fig:timeline-dataset-sizes}
%\end{figure}

% The full timeline of all $31$ $51\%$ attack events we identified is presented in Table~\ref{tab:attack-timeline-all}.
% Those events that are part of our analysis are labelled with `$E_i$', and the others with `-'.
% Note that the number $31$ is significantly more than the $14$ events reported and analysed in~\cite{SSVK2019}.
% Our list is not exhaustive.
% Given the thousands of cryptocurrencies\footnote{A list is available at~\url{https://coinmarketcap.com/all/views/all/}.} that exist today, it is possible that some $51\%$ attacks have not been reported in the media or have not appeared on Twitter, and hence have been left out from our list.
% We do not see any way to create an exhaustive list of $51\%$ attacks or threats.
Table~\ref{tab:cryptocurrencies-attacked-number-of-times} lists the cryptocurrencies and the number of times they appear in our full timeline.
Ethereum Classic tops the list with $4$ attacks.
\begin{table}[!htb]
    \centering
    \scriptsize
    \caption{The number of attacks per cryptocurrency as recorded in Table~\ref{tab:attack-timeline-all}.}
    \label{tab:cryptocurrencies-attacked-number-of-times}
    \begin{tabular}{llc}
    \hline
       Sl.~No. & Cryptocurrency  & Number of attacks \\
    \hline
       01 & Ethereum Classic & $4$ \\
       02 & Bitcoin Gold & $3$ \\
       03 & Bitcoin SV & $2$ \\
       04 & Litecoin Cash & $2$ \\
       05 & Verge & $2$ \\
       06 & Vertcoin & $2$ \\
       07 & Bitcoin$^*$ & $1$ \\
       08 & Feathercoin$^*$  & $1$ \\
       \hline
       09 & Bitcoin Private & $1$ \\
       10 & Electroneum & $1$ \\
       11 & Expanse & $1$ \\
       12 & Firo & $1$ \\
       13 & Karbo & $1$ \\
       14 & Krypton & $1$ \\
       15 & MonaCoin & $1$ \\
       16 & Pigeoncoin & $1$ \\
       17 & Powercoin & $1$ \\
       18 & Shift & $1$ \\
       19 & Terracoin & $1$ \\
       20 & ZenCash & $1$ \\
    \hline
    \end{tabular}
    \\ $^*$ Single events included in our analysis.
\end{table}

\subsection{Comparing sentiment profiles\label{sec:results-sentiments}}

\begin{figure}[!htb]
\centerline{\includegraphics[width=9cm]{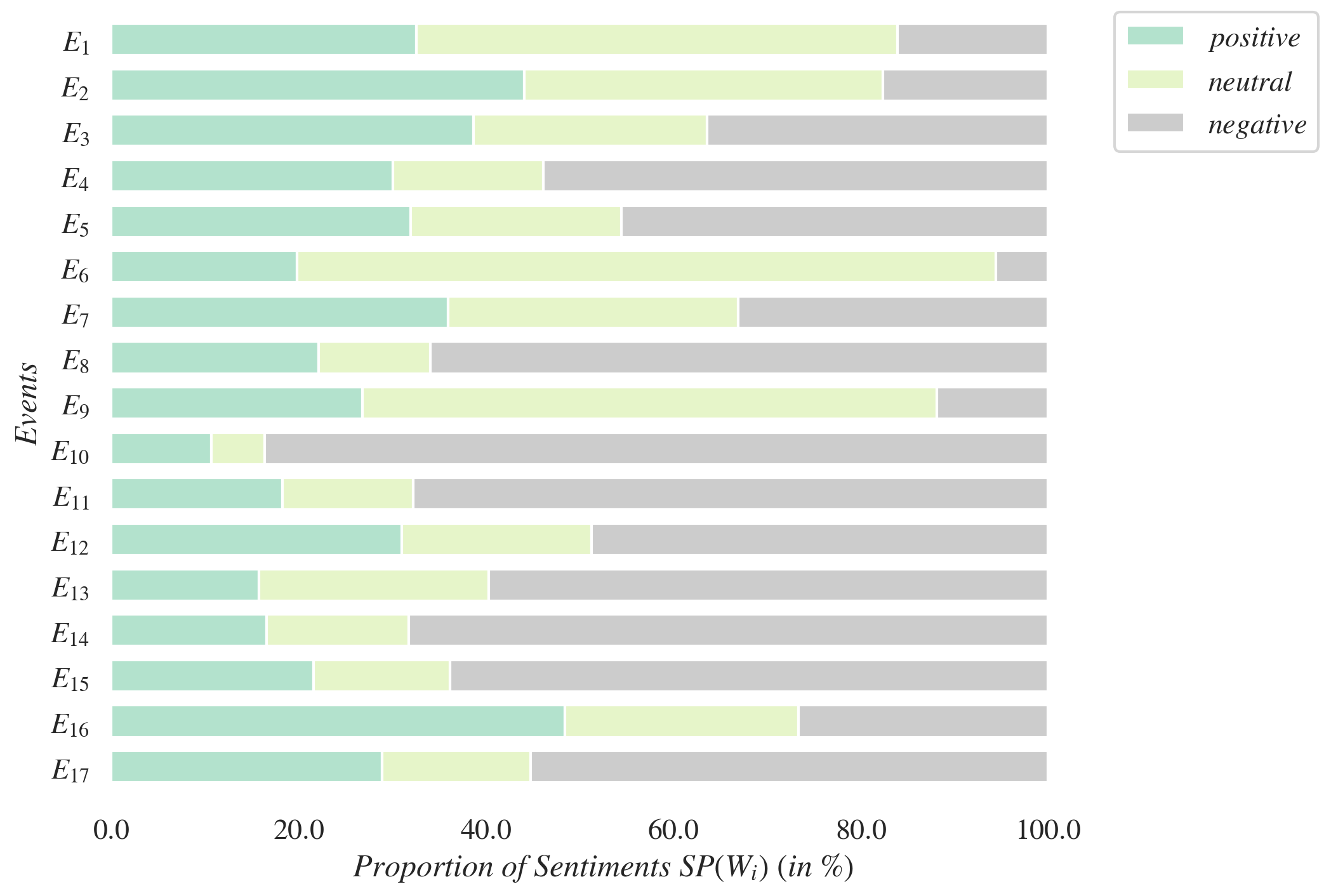}}
\caption{$\SP(\DSW_i)$: Sentiment profiles of the whole datasets $\DSW_i$}
\label{fig:sentiment-profiles-of-DSW}
\end{figure}

\begin{figure}[!htb]
\centerline{\includegraphics[width=9cm]{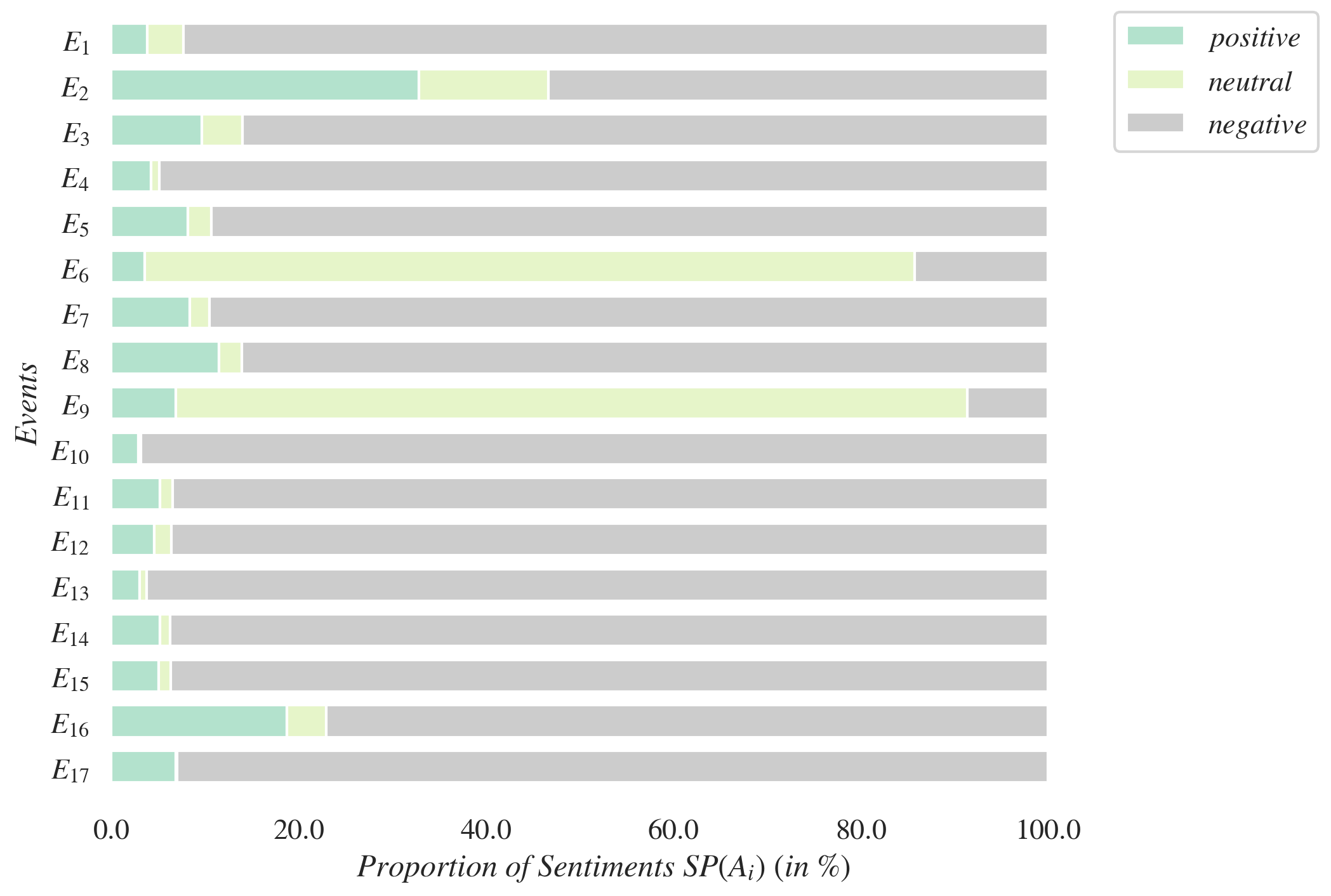}}
\caption{$\SP(\DSA_i)$: Sentiment profiles of attack datasets $\DSA_i$}
\label{fig:sentiment-profiles-of-DSA}
\end{figure}

\begin{figure}[!htb]
\centerline{\includegraphics[width=9cm] {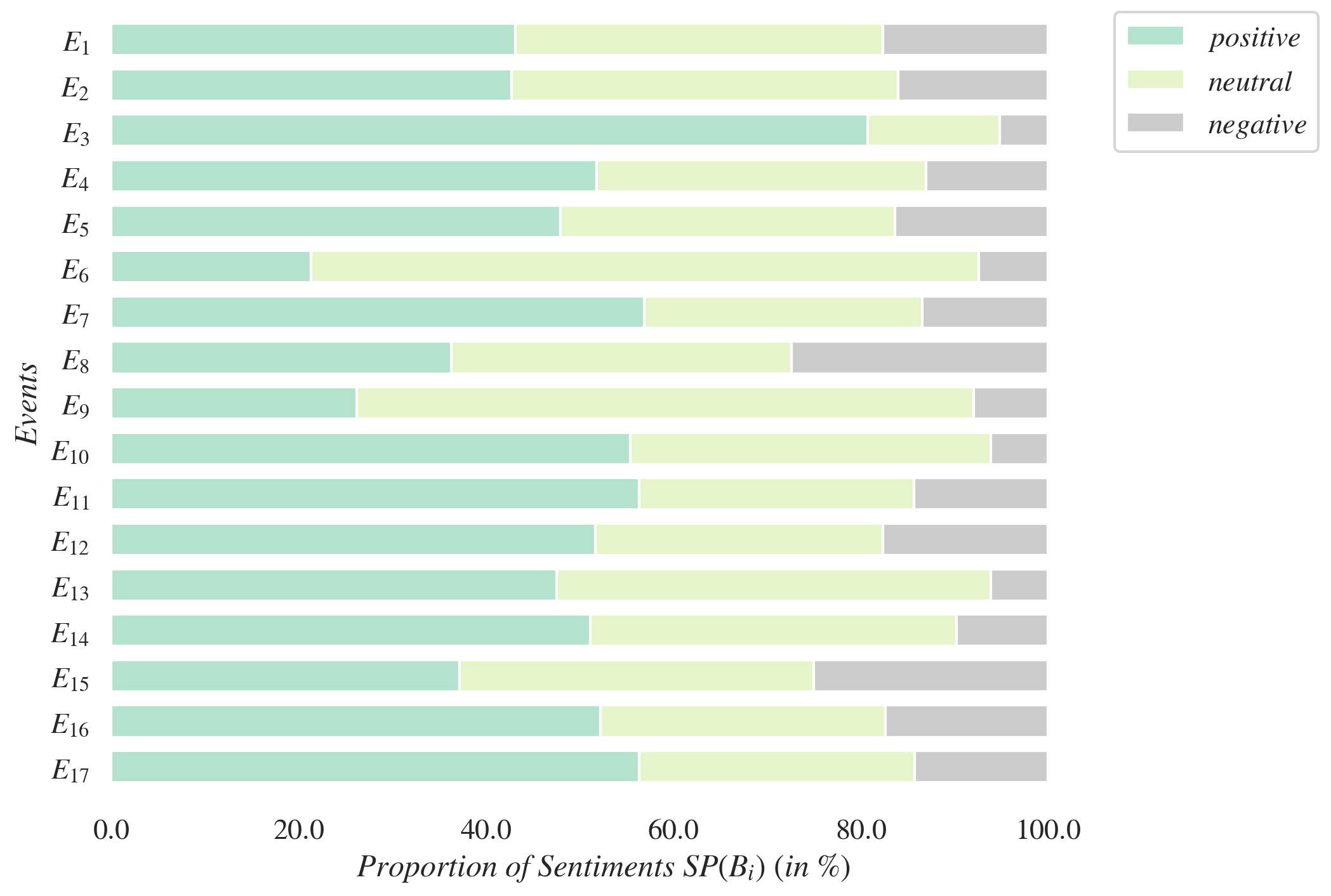}}
\caption{$\SP(\DSB_i)$: Sentiment profiles of the benchmark datasets $\DSB_i$}
\label{fig:sentiment-profiles-of-DSB}
\end{figure}

The sentiment profile $\SP(\DS)$ of a datasets $\DS$ captures the percentages of tweets categorised into positive, neutral and negative sentiments.
% We compare the events $E_i \in T$ based on the sentiment profiles of the datasets.
Figure~\ref{fig:sentiment-profiles-of-DSW}, Figure~\ref{fig:sentiment-profiles-of-DSA} and Figure~\ref{fig:sentiment-profiles-of-DSB}. %(in Appendix~\ref{sec:additional-figures})
present the sentiment profiles $\SP(\DS)$ of the datasets $\DS = \DSW_i$, $\DSB_i$ and $\DSA_i$ respectively, for all $E_i \in T$.
The sentiment profiles of the benchmark datasets $\SP(\DSB_{i})$ in Figure~\ref{fig:sentiment-profiles-of-DSB} are much more positive and have fewer negative tweets in them, when compared with $\SP(\DSW_i)$ in Figure~\ref{fig:sentiment-profiles-of-DSW} or $\SP(\DSA_i)$ in Figure~\ref{fig:sentiment-profiles-of-DSA}.
So the $51\%$ attacks are significantly noticeable events on Twitter with \textit{a clear change in the proportions of sentiments from positive/neutral to negative} for most cryptocurrencies.
This potentially indicates that sentiment profiles could be used for \textit{the effective detection of deviation from the ``norm''} and creates the possibility of designing a \textit{triggering mechanism for flagging such events}.
Such a mechanism would complement and strengthen the methodology used in this work in identifying $51\%$ attack events.

We additionally observe a clear difference in the sentiment profiles of the whole datasets $\DSW_i$ and attack datasets $\DSA_i$.
In most cases the proportion of negative tweets are significantly higher in $\DSA_i$ than in $\DSW_i$, while positive tweets are nearly eliminated from $\DSA_i$.
For example, the positive sentiments of the first known attack event $E_{1}$ on Feathercoin [8 June 2013] dropped from $32.57\%$ in $\DSW_1$ to $3.85\%$ in $\DSA_1$.
\begin{comment}
For $E_{5}$ on Verge [21-22 May 2018], the negative sentiments have increased from $45.56\%$ in $\DSW_5$ (bear in mind that $\DSW_i$ includes tweets that explicitly talk about the attack and those that do not, in the given time frame) to $89.34\%$ in $\DSA_5$.
In the case of $E_{14}$, the second attack on Ethereum Classic [29 July - 1 August 2020], while comparing $\SP(\DSW_{14})$ with $\SP(\DSA_{14})$, the positive tweets dropped from $15.79\%$ to $3\%$ while the negative sentiments increased from $59.73\%$ to $92.22\%$.
\end{comment}
The neutral tweets generally follow the same pattern of decreasing in $\DSA_i$ from $\DSW_{i}$.
However, we found two cases (both for Litecoin Cash) where the positive sentiments dropped, but the main sentiment was neutral and not negative contrary to all other events.
This could indicate that users are generally very positive about the cryptocurrency and are unfazed by $51\%$ attacks.
For the event $E_2$ of Bitcoin during 12-13 June 2014, the positive sentiments are significantly higher than in any other event.
Compared to $\DSW_2$ the positive sentiments only dropped by $11.23\%$ in $\DSA_2$ (from $44.08\%$ to $32.85\%$). 
Out of the $3409$ tweets in $\DSA_2$, there were $1120$ tweets with positive sentiments.
This could be primarily because this was only a threat and not an actual attack.
It may also indicate the high level of trust and support in the cryptocurrency community where users are positively passionate about the first and most popular cryptocurrency Bitcoin.
% Something similar, but not as intense could be observed on the first attack for $E_{18}$ on Bitcoin SV [24 June – 9 July 2021].
% \textcolor{red}{Zsofia: Perhaps, -if we not using correlations in this paper- we could still just calculate maybe pairwise t-test between the sentiments at attack and non-attack periods, and report statistically significant differences??}

\subsection{Comparing emotion profiles\label{sec:results-emotions}}

The emotion profile $\EP(\DS)$ for a dataset $\DS$ is characterised by five emotions - happiness, anger, surprise, sadness and fear.
It contains the emotion volume $\EV(\DS)$ denoting the numbers of tweets in $\DS$ that carried each of the five emotions, and the emotion intensity $\EI(\DS)$ denoting the average intensities of each of the five emotions.

\begin{figure}[!htb]
\centerline{\includegraphics[width=9cm]{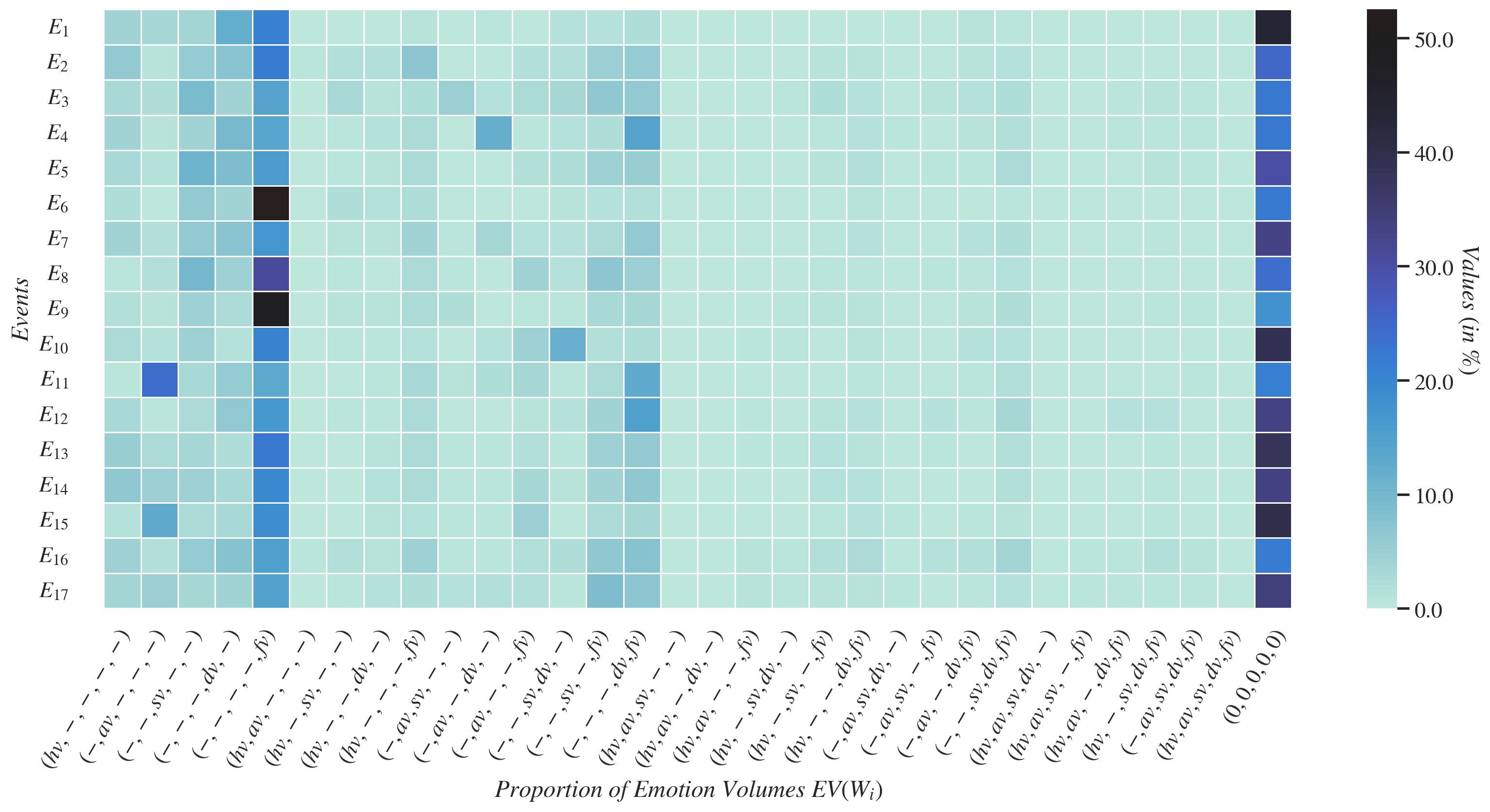}}
\caption{Heat map of percentages of tweets carrying different emotions and their combinations in $\DSW_i$ datasets}
\label{fig:emotion-volume-DSW}
\end{figure}

\begin{figure}[!htb]
\centerline{\includegraphics[width=9cm]{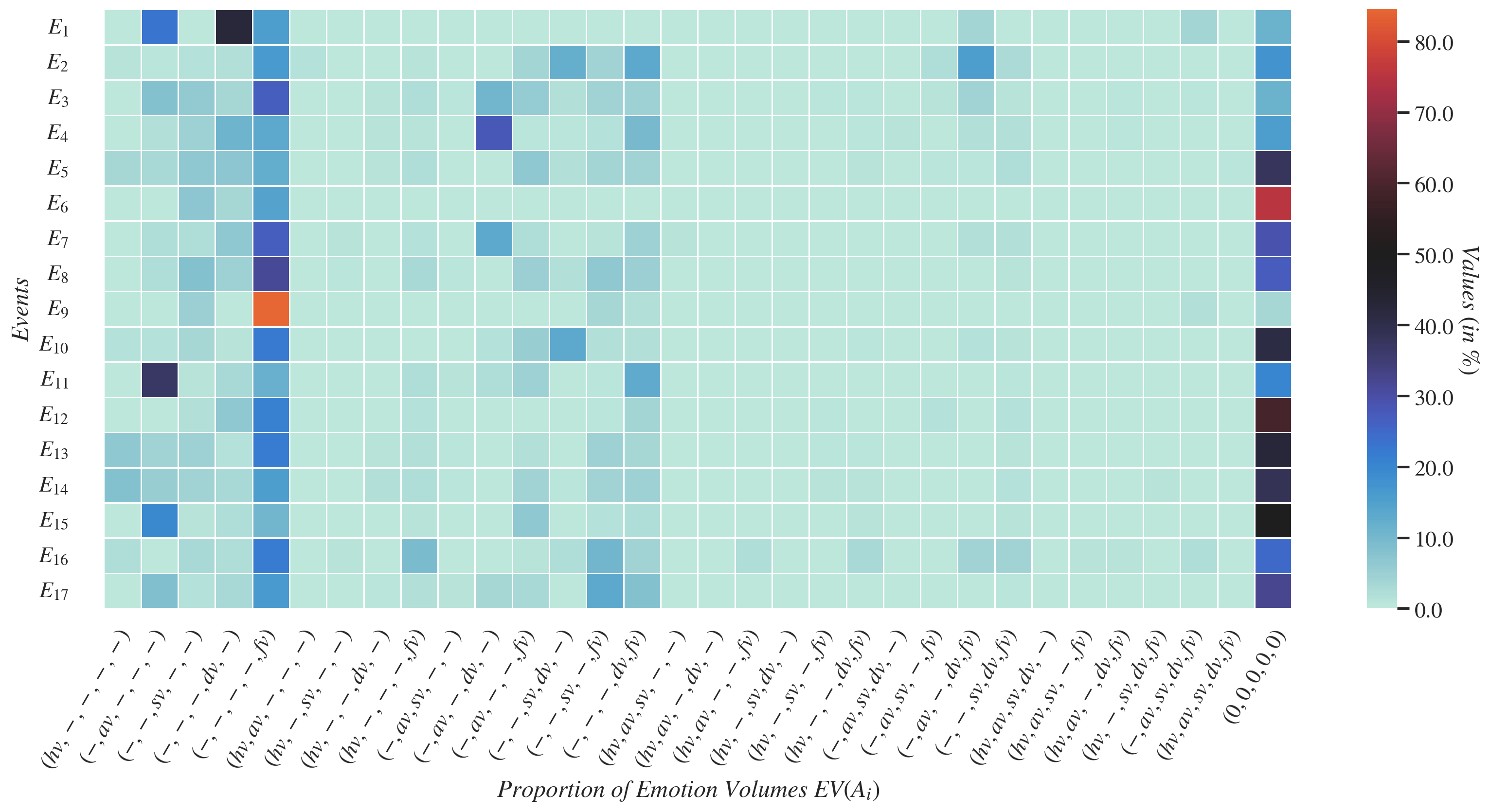}}
\caption{Heat map of percentages of tweets carrying different emotions and their combinations in $\DSA_i$ datasets}
\label{fig:emotion-volume-DSA}
\end{figure}

\begin{figure}[!htb]
\centerline{\includegraphics[width=9cm]{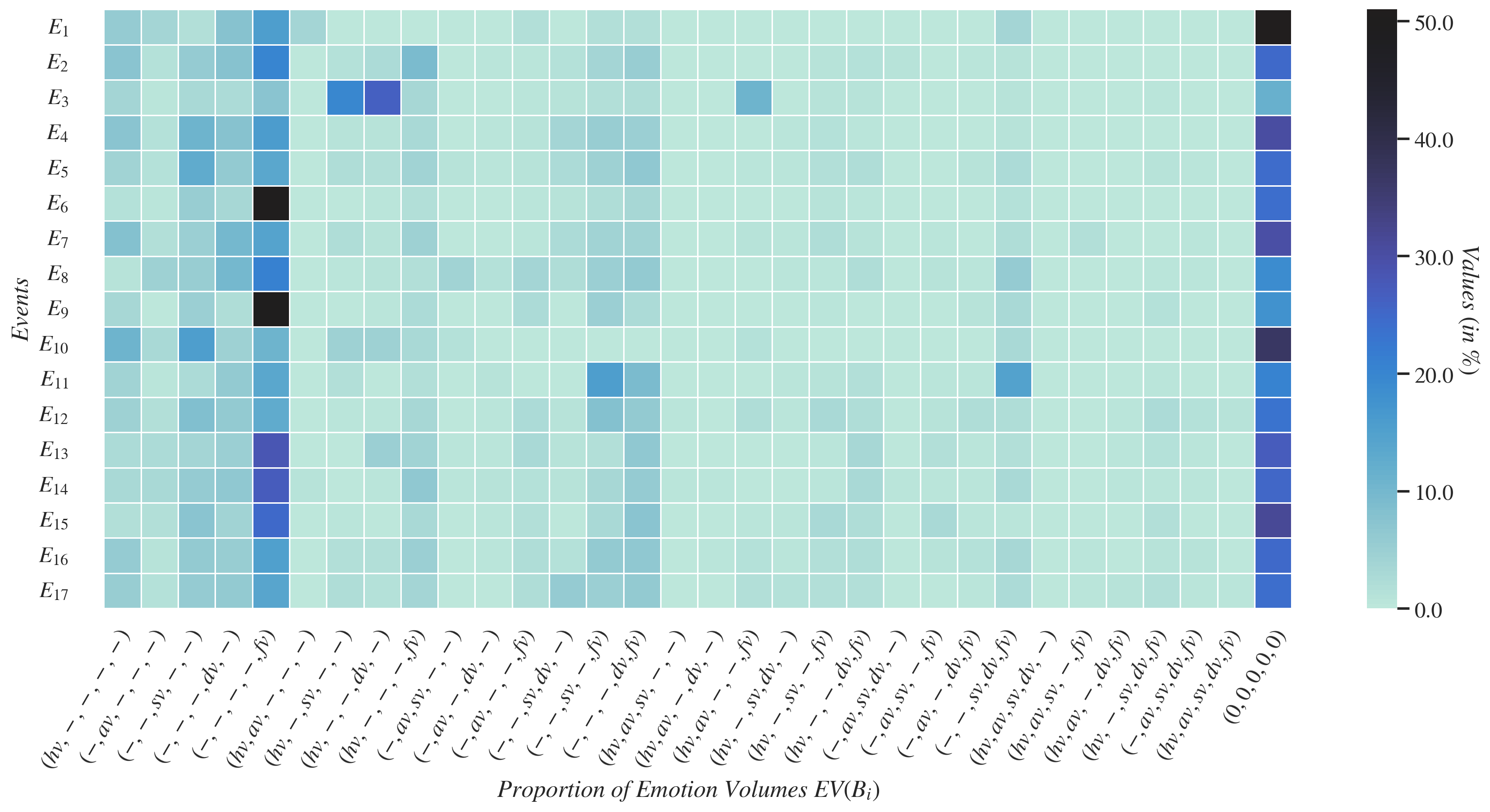}}
\caption{Heat map of percentages of tweets carrying different emotions and their combinations in $\DSB_i$ datasets}
\label{fig:emotion-volume-DSB}
\end{figure}

The emotion volumes $\EV(\DSW_i), \EV(\DSB_i)$ and $\EV(\DSA_i)$ for the respective datasets are represented as heat maps in Figures~\ref{fig:emotion-volume-DSW},~\ref{fig:emotion-volume-DSA} and~\ref{fig:emotion-volume-DSB}
% (in Appendix~\ref{sec:additional-figures})
respectively.
The rows in a heat map corresponds to the events $E_i \in T$.
The columns are for individual emotions or their combinations.
The rightmost column labelled $(0, 0, 0, 0, 0)$ is for \textit{neutral} tweets in which none of the five emotions were detected.
For any other column, its label indicates the emotions counted therein.
For example, the first column labelled $(HV, -, -, -, -)$ is for tweets in which only the happy emotion was detected.
The column labelled $(-, -, SV, -, FV)$ is for tweets that carry only sadness and fear.
A cell in the heat map is for the tweets that carry only the emotions indicated in its column label, for the event $E_i$ denoted in its row label, in the dataset denoted in the heat map figure caption.
The colour of the cell denotes the percentage of tweets in the dataset ($\DSW_i, \DSB_i$ or $\DSA_i$) corresponding to the figure, as indicated in its legend to its right.
Darker colours denote higher percentages.
A tweet gets counted in a cell if the emotions in its column label have been detected in it, regardless of the magnitudes of the emotions in the tweet.
The row for $E_i$ in Figure~\ref{fig:emotion-volume-DSW} (respectively Figures~\ref{fig:emotion-volume-DSB} and~\ref{fig:emotion-volume-DSA}) is essentially a partition of the total count of tweets in the dataset $\DSW_i$ (respectively $\DSB_i$ and $\DSA_i$) into the counts of tweets carrying the specific emotions indicated in the respective column labels.
So the counts in the cells of each row add up to $100\%$.

Our observations from the heat maps are as follows.
For majority of the tweets in almost every $E_i \in T$, no emotions have been detected and hence they have been counted in the last column labelled $(0, 0, 0, 0, 0)$.
This is evident from the rightmost column being the darkest in each of the Figures~\ref{fig:emotion-volume-DSW},~\ref{fig:emotion-volume-DSB} and~\ref{fig:emotion-volume-DSA}.
However, when emotions are expressed, they are mostly one at a time in a tweet.
Hence the first five columns representing tweets with singleton emotions are the next darkest ones.
Among these first five columns, the fifth column for fear is usually darkest, indicating it to be the most dominant singleton emotion.
The fifth column is also generally darker in Figure~\ref{fig:emotion-volume-DSW} than in Figure~\ref{fig:emotion-volume-DSB}, denoting that \textit{fear grows at the time of a $51\%$ attack on a cryptocurrency, compared to when it is not being attacked}.
When emotions occur in pairs or when more than two emotions are present in a tweet, fear is also usually one of them.
In the attack datasets $\DSA_i$ (Figure \ref{fig:emotion-volume-DSA}), some interesting patterns could be observed that differ from the whole datasets $\DSW_i$.
Generally, the emotions are more intensely present in $\DSA_i$.
Fear is the most dominant emotion in most $\DSW_i$, while there are some significant shifts toward anger, sadness and neutrality in $\DSA_i$.
In benchmark datasets $\DSB_i$ (Figure~\ref{fig:emotion-volume-DSB}), happiness both as a singleton and in various combinations with the other emotions, is more significantly present, while anger is hardly present in them.
So, \textit{people are generally happy with cryptocurrencies, until they are attacked}.

\begin{comment}
We can see examples where the most dominant emotion or combination of emotions shift from the neutral $(0, 0, 0, 0, 0)$.
For example, both anger and sadness appeared together most significantly in event $E_{4}$ of Bitcoin Gold [16-19 May 2018] with $28.14\%$ or anger for event $E_{11}$ on Bitcoin Gold [23-24 January 2020] with $36.76\%$.
\end{comment}

There are some deviations from our previous observations.
For event $E_{11}$ on Bitcoin Gold [23-24 Jan 2020], within the whole dataset $\DSW_{11}$, the most dominant emotion expressed is anger, followed by no emotions in neutral tweets and then the pair (sadness, fear).
It is also striking that Litecoin Cash whose sentiment profile in $\DSA_i$ previously showed that it mainly has neutral sentiments, still has $75\%$ of neutral tweets with no emotions in the first instance ($\DSA_{6}$), whereas during the second attack ($\DSA_{9}$), fear is overwhelmingly dominant at $84.48\%$.
\textit{This demonstrates the advantage of the granularity in capturing reactions through the emotion profile over the sentiment profile.}

\begin{comment}
In the further analysis of emotions we summarised the above categories, and looked at all the tweets that contain either of the five emotions. 
Note that there are overlaps, one tweet could fall into more categories.
For example any tweet that contains more than one emotion will count towards all those emotion categories.
The main interest with this approach is to summarise that each of the five emotions how often appears in the tweets. 
\end{comment}

\begin{figure}[!htb]
\centerline{\includegraphics[width=9cm]{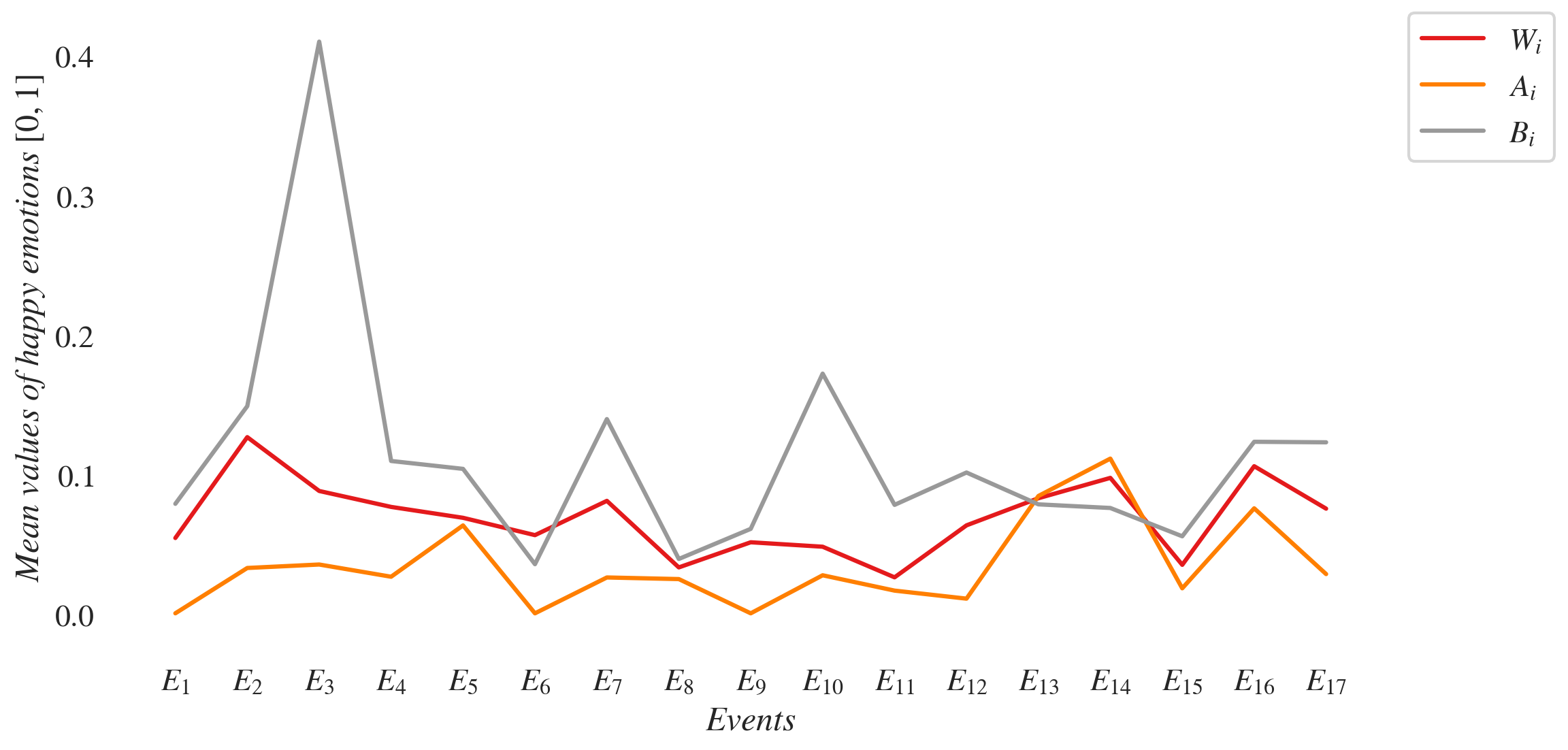}}
\caption{$H(\DSW_i), H(\DSA_i), H(\DSB_i)$: Mean intensities of happiness across the datasets $\DSW_i, \DSA_i, \DSB_i$ for all events $E_i \in T$}
\label{fig:intensities-happiness}
\end{figure}

\begin{figure}[!htb]
\centerline{\includegraphics[width=9cm]{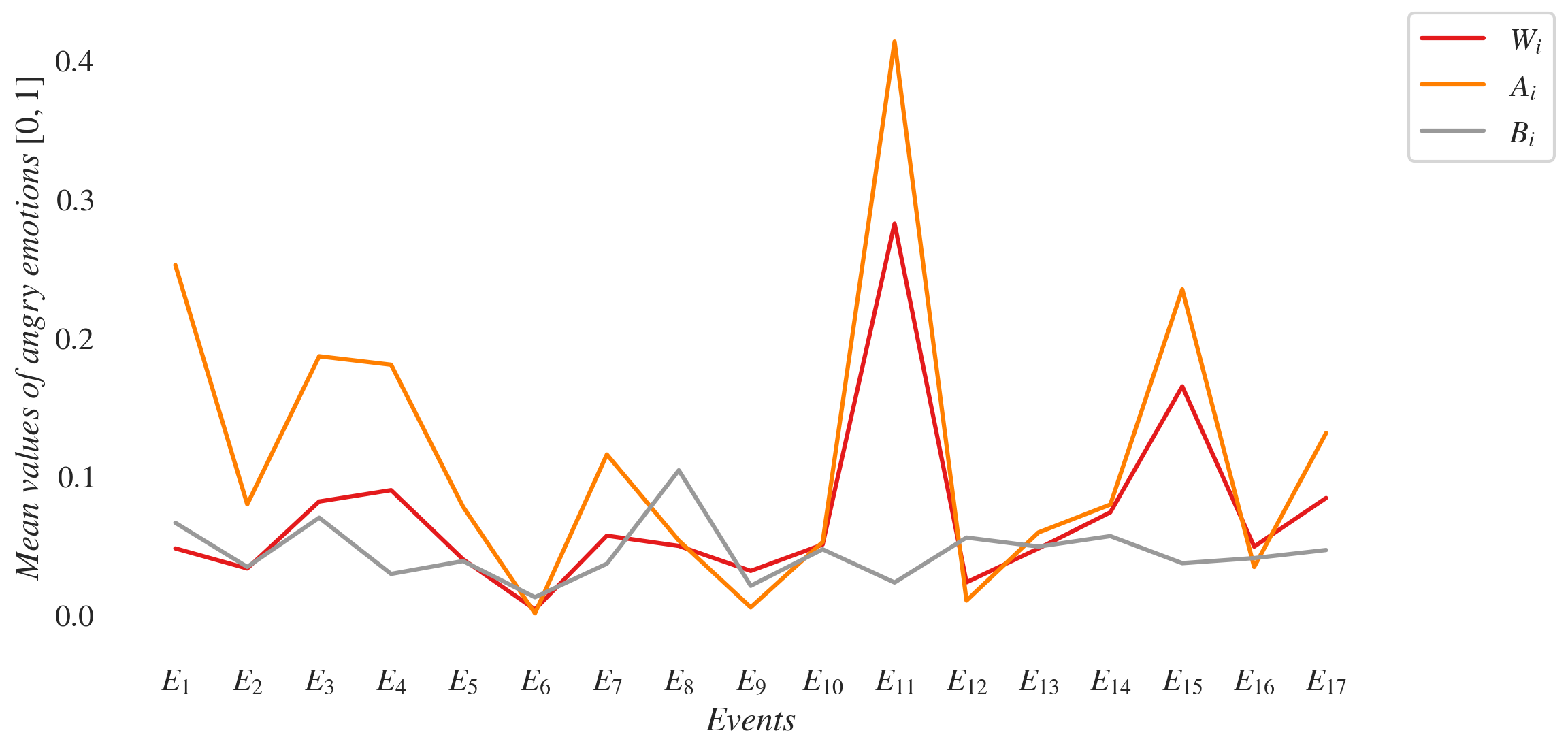}}
\caption{$A(\DSW_i), A(\DSA_i), A(\DSB_i)$: Mean intensities of anger across the datasets $\DSW_i, \DSA_i, \DSB_i$ for all events $E_i \in T$}
\label{fig:intensities-anger}
\end{figure}

\begin{figure}[!htb]
\centerline{\includegraphics[width=9cm]{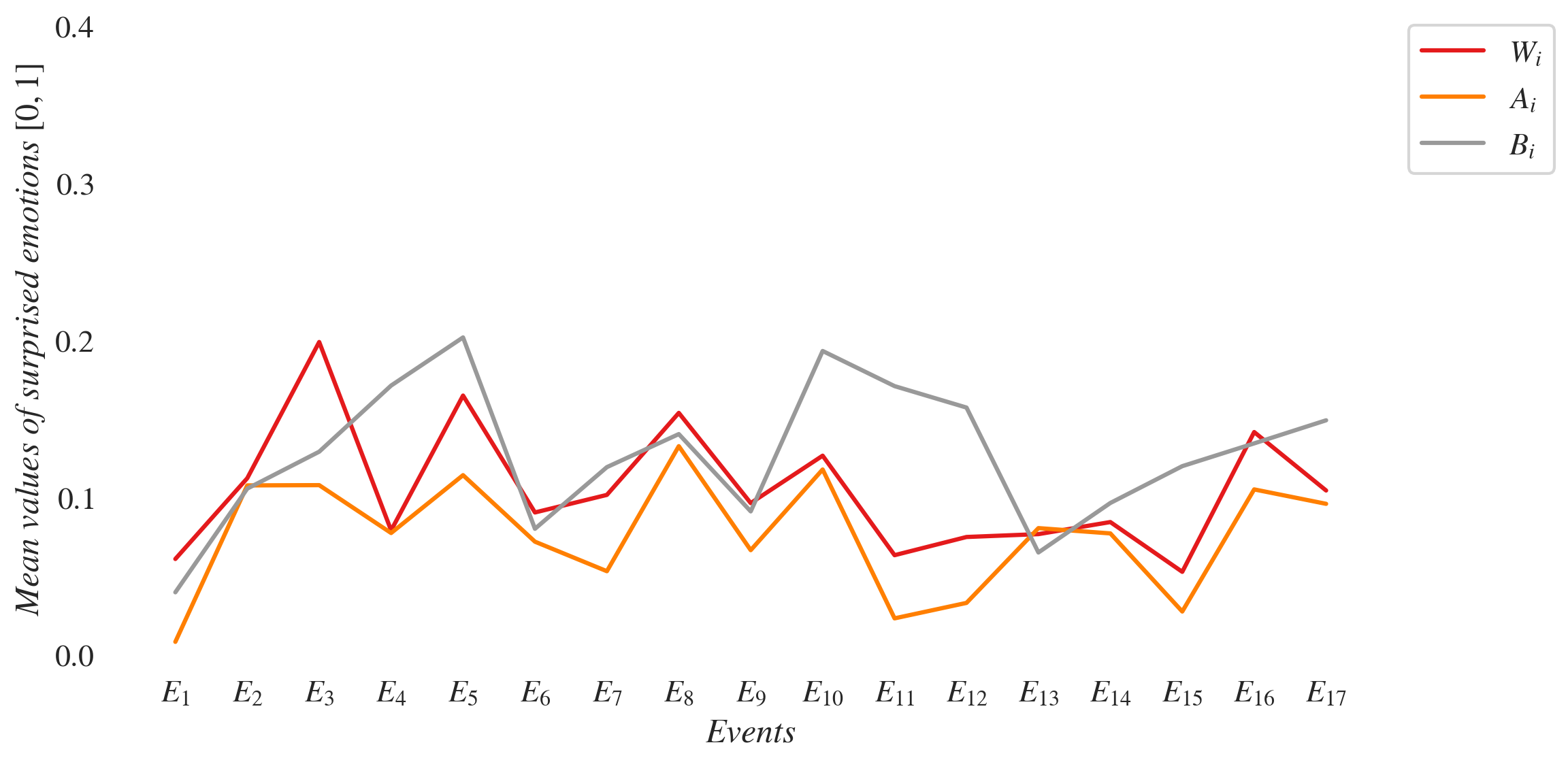}}
\caption{$S(\DSW_i), S(\DSA_i), S(\DSB_i)$: Mean intensities of surprise across the datasets $\DSW_i, \DSA_i, \DSB_i$ for all events $E_i \in T$}
\label{fig:intensities-surprise}
\end{figure}

\begin{figure}[!htb]
\centerline{\includegraphics[width=9cm]{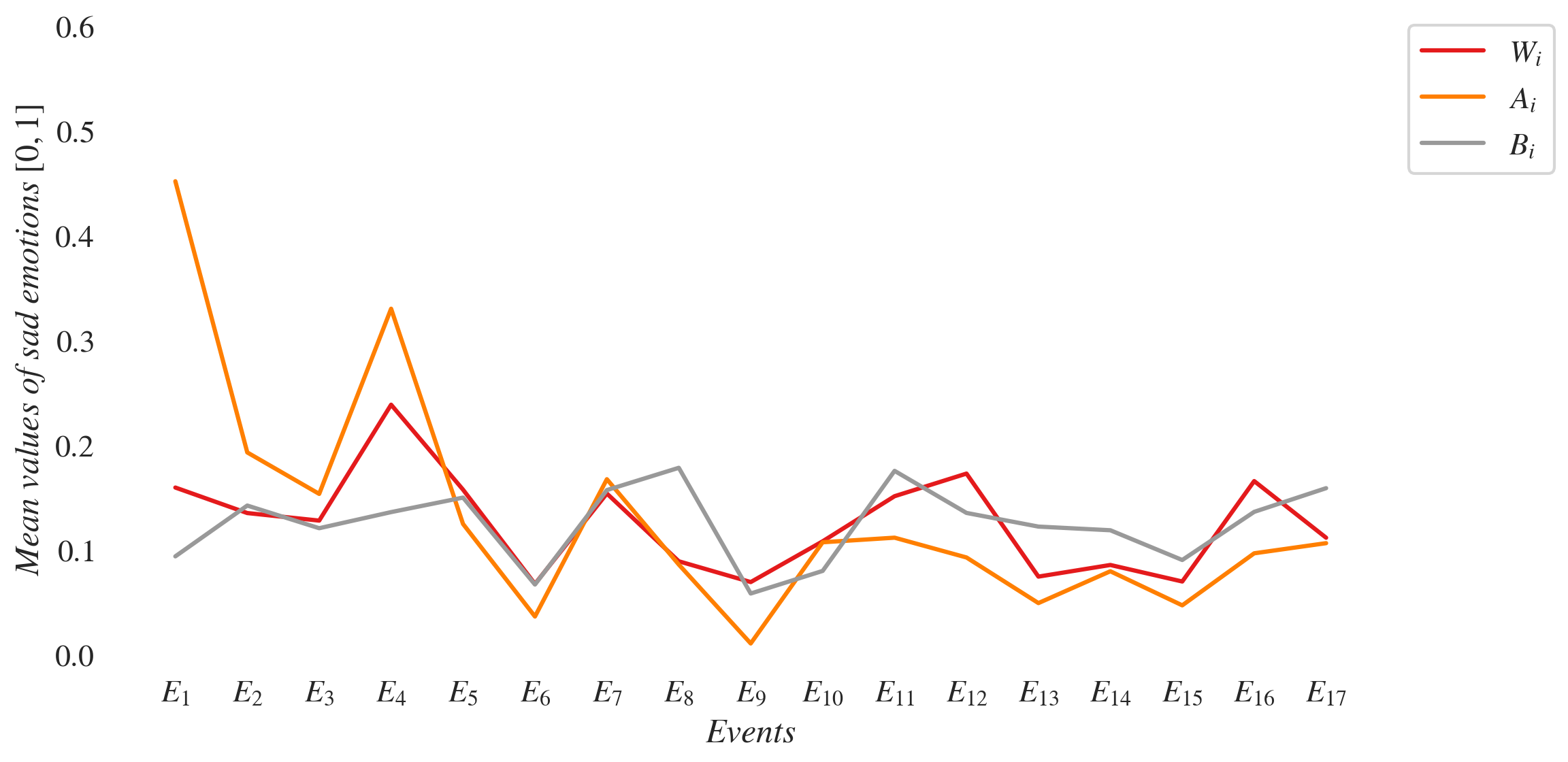}}
\caption{$D(\DSW_i), D(\DSA_i), D(\DSB_i)$: Mean intensities of sadness across the datasets $\DSW_i, \DSA_i, \DSB_i$ for all events $E_i \in T$}
\label{fig:intensities-sadness}
\end{figure}

\begin{figure}[!htb]
\centerline{\includegraphics[width=9cm]{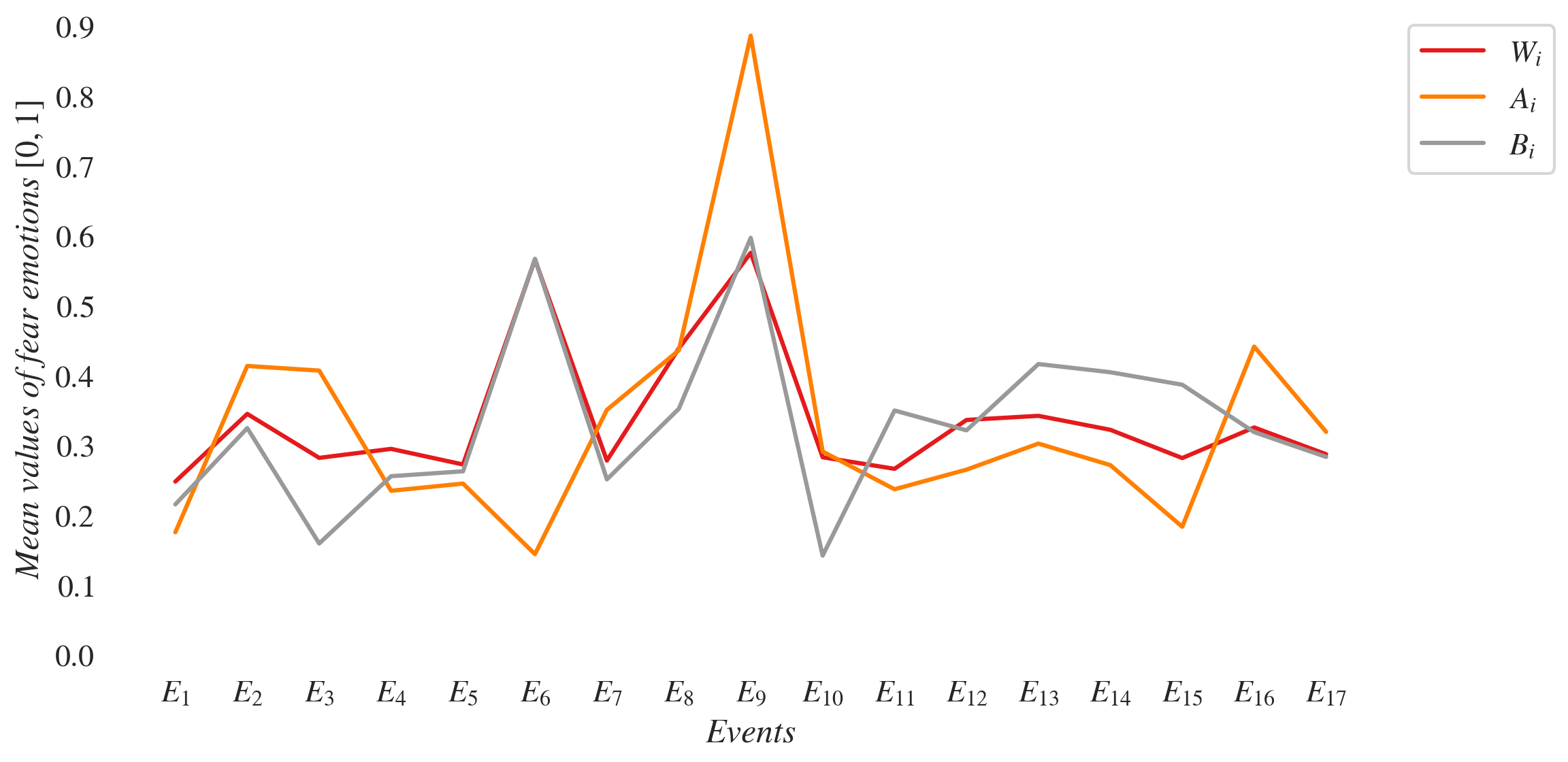}}
\caption{$F(\DSW_i), F(\DSA_i), F(\DSB_i)$: Mean intensities of fear across the datasets $\DSW_i, \DSA_i, \DSB_i$ for all events $E_i \in T$}
\label{fig:intensities-fear}
\end{figure}

We next consider the emotion intensities for the datasets $\DSW_i, \DSB_i$ and $\DSA_i$.
Recollect that the emotion intensity of the dataset $\DS$ is given by
\[\EI(\DS) = (H(\DS), A(\DS), S(\DS), D(\DS), F(\DS)).\]
These mean intensities of happiness, anger, surprise, sadness and fear have been plotted in Figures~\ref{fig:intensities-happiness},~\ref{fig:intensities-anger},~\ref{fig:intensities-surprise},~\ref{fig:intensities-sadness} and~\ref{fig:intensities-fear} respectively.
% \red{Figures~\ref{fig:intensities-anger},~\ref{fig:intensities-surprise} and~\ref{fig:intensities-sadness} are in Appendix~\ref{sec:additional-figures}.}

We observe the following on emotion intensities.
As expected, anger, sadness and fear are more intense in the attack datasets $\DSA_i$ (orange lines);
happiness is generally more intense in the benchmark datasets $\DSB_i$ (grey lines).
The plot for happiness (Figure~\ref{fig:intensities-happiness}) shows a good level of consistency.
In almost all attack datasets $\DSA_i$ their means are lower than those in $\DSW_i$ and $\DSB_i$ with only the exceptions of the second and third attack on Ethereum Classic (events $E_{13}$ and $E_{14}$),
while anger (Figure~\ref{fig:intensities-anger}) is more intense.
The intensity of surprise (Figure \ref{fig:intensities-surprise}) generally averages below $0.2$ across all datasets.
There is of course a noticeable high peak at $E_2$ (the Bitcoin event) in the whole dataset (red line).
This particular event did shake the community and hence the element of surprise.
However, not all emotions across the datasets show a consistent pattern.
Sadness (Figure~\ref{fig:intensities-sadness}) and fear (Figure~\ref{fig:intensities-fear}) have generally varied across datasets.
Approximately half of the time in $\DSA_i$, their intensities are either under or above compared to $\DSW_i$ and $\DSB_i$.
These two emotions therefore do not give a consistent picture across cryptocurrencies.

\textit{Fear has the highest mean intensity across all datasets}, compared to the other four emotions.
It fluctuates roughly between $0.2$ and $0.9$, while the others have their means around $0.1$.
Studying fear closely for the various cryptocurrencies, we see that Bitcoin, Vertcoin and Bitcoin SV have their mean intensities for $\DSA_i$ above $\DSW_i$ and $\DSB_i$.
On the other hand, Feathercoin, Bitcoin Gold and Ethereum Classic have their mean intensities for $\DSA_i$ slightly under $\DSW_i$ and $\DSB_i$.
The third cluster is of Verge and Litecoin Cash that have fluctuating intensities in subsequent attacks.
It is also interesting to note that in case of the second attack on Litecoin Cash ($E_{9}$) there is a high peak fear intensity of around $0.9$ (the maximum score could only be $1$).
Note that the sentiment profile had only shown that a vast majority of the tweets are neutral.
\textit{This further demonstrates that the more granular emotion profile provides better insights than the sentiment profile.}
Moreover, these currency-specific observations could indicate \textit{behavioural differences between the various cryptocurrency communities and their supporters}.

\subsection{Analysis of peak days\label{sec:results-peakdays}} 

\begin{figure}[!htb]
\centerline{\includegraphics[width=9cm]{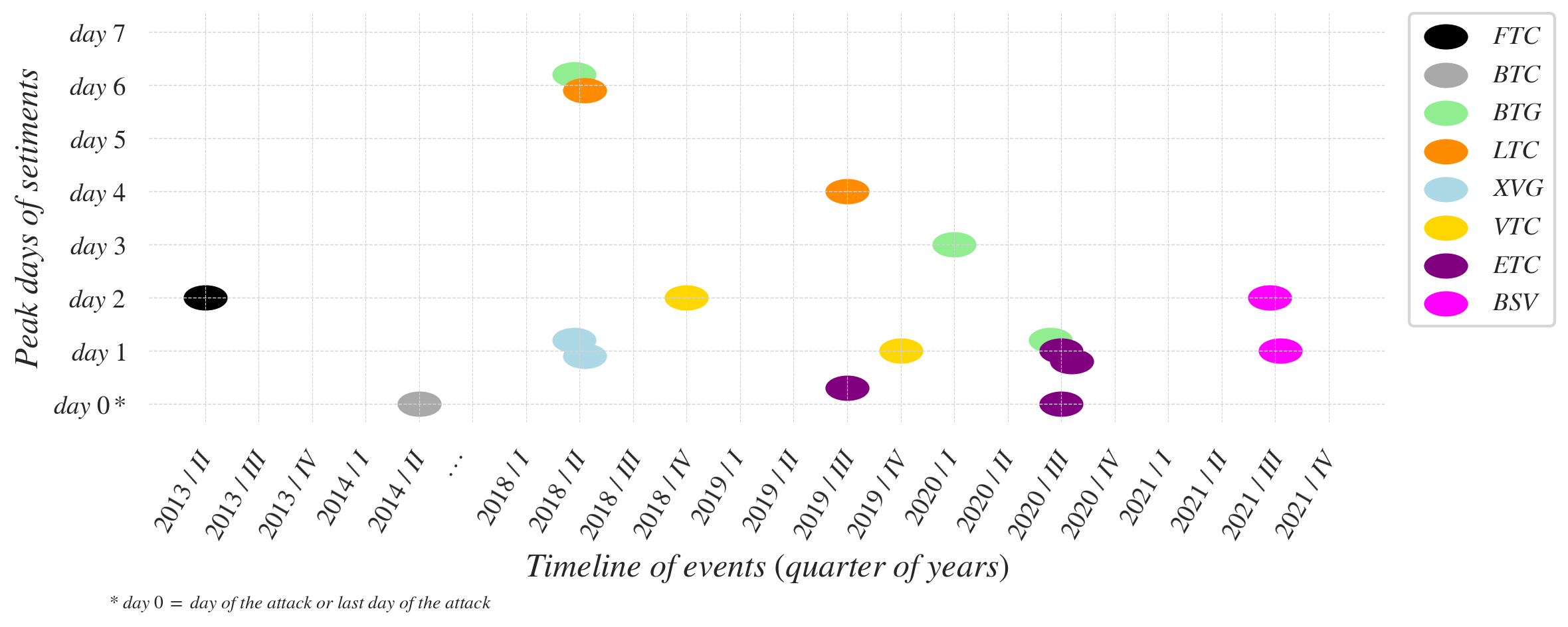}}
\caption{Peak day delays}
\label{fig:peak-day-delays}
\end{figure}

We further analyse the events $E_i \in T$ by identifying the peak days in terms of the number of tweets, following its attack period $(t^s, t^e)$.
We count the peak day starting from the last day of the attack $t^e$ as day $0$.
Table \ref{tab:peak-days-with-dominant-sentiments} provides the peak days for each $E_i \in T$ and the dominant sentiments in the respective $\DSA_i$.
Figure~\ref{fig:peak-day-delays} denotes the distances of the peak days from day $0$.
Here, the timeline is presented as quarter of year blocks and the cryptocurrencies are colour coded and placed in the timeline.

We observe that the sentiments on the peak days are in most cases overwhelmingly negative.
Interestingly, \textit{there is a reaction delay in attaining the peak} from day $0$.
In most cases, the peak is delayed by a day or two.
Sometimes a quick same day reaction (as identified through the time in the tweet) has the same delay as one from the next day because of variations in time-zones.
There could be other factors like the time of the day when the attack is first identified and reported.
If the information came out in the late hours of the day anywhere on earth, it would fall into the next day category.
This anomaly could be observed for example in the case of attacks on more popular cryptocurrencies like Ethereum Classic which have enough data for the anomaly to be identified and distinguished.
\textit{In majority of the cases, the reaction delays were initially higher, but significantly decreased in subsequent attacks.}
For example, in the first attack on Bitcoin Gold (BTG), the peak was on the sixth day after the attack ended on 19 May 2018.
However, for the second attack, the peak was on the third day after 24 January 2020; for the third attempted attack, the peak was on just the next day after 10 July 2020.
Another example is Litecoin Cash (LTC) where the peak was attained after the sixth day for the first attack in 2018, while in 2019 it decreased to the fourth day.

\begin{table}[!htb]
\scriptsize
\caption{Peak days of post-attack tweet volume and their dominant sentiments}
\begin{center}
\begin{tabular}{lllll}
\hline 
$E_i$ & \textbf{Currency}  & \textbf{Attack period} & \textbf{Peak day} & \textbf{Main sentiment} \\ % & \textbf{Main emotion} \\
\hline
$E_{1}$  & Feathercoin        & (08 Jun 2013, 08 Jun 2013) & 10 Jun    & negative \\ % & Sad\\
$E_{2}$  & Bitcoin            & (12 Jun 2014, 13 Jun 2014) & 16 Jun    & negative \\ % & Fear \\
$E_{3}$  & Verge              & (04 Apr 2018, 06 Apr 2018) & 05 Apr    & negative \\ % & Fear \\
$E_{4}$  & Bitcoin Gold       & (16 May 2018, 19 May 2018) & 24 May    & negative \\ % & Sad \\
$E_{5}$  & Verge              & (21 May 2018, 22 May 2018) & 23 May    & negative \\ % & Neutral \\
$E_{6}$  & Litecoin Cash      & (30 May 2018, 30 May 2018) & 5 June    & neutral  \\ % & Neutral \\
$E_{7}$  & Vertcoin           & (02 Dec 2018, 02 Dec 2018) & 04 Dec    & negative \\ % & Fear \\
$E_{8}$  & Ethereum Classic   & (05 Jan 2019, 08 Jan 2019) & 08 Jan    & negative \\ % & Fear \\
$E_{9}$  & Litecoin Cash      & (04 Jul 2019, 07 Jul 2019) & 11 Jul    & neutral  \\ % & Fear\\
$E_{10}$ & Vertcoin           & (01 Dec 2019, 01 Dec 2019) & 02 Dec    & negative \\ % & Neutral \\
$E_{11}$ & Bitcoin Gold       & (23 Jan 2020, 24 Jan 2020) & 27 Jan    & negative \\ % & Angry \\
$E_{12}$ & Bitcoin Gold       & (02 Jul 2020, 10 Jul 2020) & 11 Jul    & negative \\ % & Neutral \\
$E_{13}$ & Ethereum Classic   & (29 Jul 2020, 01 Aug 2020) & 01 Aug    & negative \\ % & Neutral \\
$E_{14}$ & Ethereum Classic   & (05 Aug 2020, 05 Aug 2020) & 06 Aug    & negative \\ % & Neutral \\
$E_{15}$ & Ethereum Classic   & (29 Aug 2020, 29 Aug 2020) & 30 Aug    & negative \\ % & Angry \\
$E_{16}$ & Bitcoin SV         & (24 Jun 2021, 09 Jul 2021) & 11 Jul    & negative \\ % & Neutral \\
$E_{17}$ & Bitcoin SV         & (03 Aug 2021, 03 Aug 2021) & 04 Aug    & negative \\ % & Fear \\
\hline
\end{tabular}
\label{tab:peak-days-with-dominant-sentiments}
\end{center}
\end{table}

\section{Conclusion}
\label{sec:conclusion}
Given that security against the $51\%$ attack is fundamental to the integrity of blockchain based cryptocurrencies, they deserve more attention and thorough scrutiny.
In this work, we have created a timeline of some such events on various cryptocurrencies.
We have conducted volumetric analysis of the tweets that suggest that interested cryptocurrency users are concerned about the $51\%$ attack.
We have characterised the datasets using the VADER lexicon's compound scores for positive, neutral and negative sentiments.
We observed that the sentiments are overwhelmingly negative at the time of the attack as compared to normal periods.
We have further characterised the datasets using the Text2Emotion lexicon scores of five emotions (happy, angry, surprise, sad, and fear).
We observed that fear is generally predominant at all times, for all cryptocurrencies, and becomes significantly dominant especially at the time of attacks.
Our analysis shows that the volume and perceptions of tweets could work as a triggering system to alert users of an ongoing attack.
We point out that such a system should be equipped to handle mis-/dis-information on social media to make it more reliable.
We are not aware of any such specialised system for handling mis-/dis-information on cryptocurrencies and leave the selection/development of a suitable system as future work.

%\bibliographystyle{unsrt}
%\bibliography{ccsa}

\appendix

%\section{Further details on the timeline of 51\% Attacks}
%\label{sec:further-details-on-timeline}

%\section{Additional figures}
%\label{sec:additional-figures}

\section{Background and Preliminaries\label{sec:background-andprelims}}
% \textcolor{red}{Reviewer 2, minor issues: "2.1 in the background does not contribute much to understanding the rest of the paper, as none of the technical contributions has to do with forks, the storage model of the blockchain chain, or forking sustaining."}

\subsection{Blockchain and cryptocurrency background\label{sec:crypto-background}}
Public blockchains are decentralised and distributed systems.
As a data structure, a blockchain is a linear collection of blocks to which new ones can be added at one end.
We denote a block as $B_i$ indexed by its position $i$ in the chain
\[\BC = \left(B_0, B_1, \ldots, B_n\right)\]
of blocks.
The first block $B_0$ is called the \textit{genesis block} and the \textit{block height} of $B_i$ is $i$.

A \textit{proof-of-work} blockchain involves a cryptographically secure hash function $H:\{0,1\}^* \to \{0,1\}^t$ in the process of finding a new block.
A block $B_{i}$ contains the hash $H(B_{i-1})$ of the previous block $B_{i-1}$ in the chain and a \textit{nonce} $\eta_i$ such that its own hash $H(B_i)$ has a pre-determined prefix (typically a certain number of $0$'s) in its $t$-bit output value.
The prefix is determined by the time that was required to generate (some of) the previous blocks in the chain.
In order to mine a valid block, a miner searches for the appropriate nonce $\eta_i$.
Finding the nonce is the computational puzzle that gets more and more difficult as the length of the required hash prefix increases.
When an appropriate nonce has been found, it is considered as a probabilistic evidence that the miner has done the estimated amount of computational work, and hence the name proof-of-work.

\subsection{Forks and $51\%$ Attacks\label{sec:forks-51attacks}}
Miners usually maintain their own local copy of the entire blockchain data structure.
A secure blockchain system promises to provide an {immutable history}.
This means that once a block $B_i$ has been added to the blockchain, the probability that it may be removed from it decreases exponentially as more blocks $B_j, j > i$ are added thereafter.

The possibility that a block that was once added to the chain may get removed from it is due to a \emph{fork} in the chain.
A fork occurs when there are two or more chains with common histories.
For example,
\[\BC = \left(B_0, B_1, \ldots, B_{i-1}, B_i, B_{i+1}, \ldots, B_n\right)\]
and
\[\BC' = \left(B_0, B_1, \ldots, B_{i-1}, B_i', B_{i+1}', \ldots, B_m'\right)\]
are two chains forked from the $i^{th}$ block such that they have the common history $B_0, B_1, \ldots, B_{i-1}$, but then the first chain has the blocks $B_i, B_{i+1}, \ldots, B_n$ while the second chain has the blocks $B_i', B_{i+1}', \ldots, B_m'$.

Forks may occur for various reasons in a blockchain.
The creation of new blocks is a randomised event.
Being a decentralised system, there is no mechanism of coordination between competing miners in their race to create distinct new blocks.
As a result, it is often the case that two or more miners create their own new blocks $B_{i+1}$ and $B_{i+1}'$ on top of the same block $B_{i}$ at almost the same time.
However, the blockchain protocol dictates that \textit{the most difficult (or the longest) chain be considered as valid} at any point of time.
This means that only one of these two blocks $B_{i+1}$ and $B_{i+1}'$ makes it to the chain that would be considered valid in the long run.
The other block gets \textit{orphaned}.
It is possible that the blockchain network gets divided into two parts such that one set of miners attempts to create blocks on top of $B_{i+1}$ while the others attempts to create blocks on top of $B_{i+1}'$.
This results in a more sustained fork.
Assuming that all miners choose to follow only one of the two chains, they abandon mining on top of one of the chains and continue mining on the other one.
Hence all blocks after the fork in the abandoned chain get orphaned.

A sustained fork may be created with malicious intent.
This happens when a set of miners divert from the protocol of building upon the most difficult chain, and instead choose to mine blocks as a fork of the chain followed by the rest of the miners in the network.
To launch such an attack successfully on a connected network, the attacker set of miners should possess at least $50\%$ of the resources for mining new blocks.
Such an attack is called the \textit{$51\%$ attack}.

A sustained fork may occur when miners disagree on the proposed protocol changes.
A set of miners start following a new protocol while the old protocol is followed by the rest of the miners.
The blocks generated through one protocol are considered invalid by the other.

A cryptocurrency system uses a blockchain as a ledger of transactions.
A unit of the cryptocurrency is called a \textit{coin}.
Each block records transactions that either (1) create a certain number of new coins, or (2) transfer those coins from one owner to another.
Ownership of coins is ascertained by a cryptographic digital signature system.
To transfer the ownership of a coin, a sender uses their secret key to generate a digital signature on a transcript containing the recipient's public key.
This signature can be verified using the public key of the sender.
An adversary may \textit{hack} into a cryptocurrency owner's wallet (or the exchanges they use) to steal the secret keys.
Once they get hold of the secret key or they are able to forge the owner's signature by some other means, they can steal coins from the owner by generating new transactions that subsequently enter the chain.
However, this is not an attack on the underlying blockchain.
In this work, we study only $51\%$ attacks on cryptocurrency blockchains and not the hacks or other such attacks on cryptocurrencies.

\end{document}